\documentclass[twocolumn, twocolappendix]{aastex631}
\usepackage{booktabs}
\usepackage{float}

\begin{document}

\title{Dust Survival in Galactic Winds}

\author[0000-0001-6325-9317]{Helena M. Richie}
\affiliation{Department of Physics and Astronomy, University of Pittsburgh, 3941 O’Hara St, Pittsburgh, PA 15260, USA}

\author[0000-0001-9735-7484]{Evan E. Schneider}
\affiliation{Department of Physics and Astronomy, University of Pittsburgh, 3941 O’Hara St, Pittsburgh, PA 15260, USA}

\author[0000-0002-7918-3086]{Matthew W. Abruzzo}
\affiliation{Department of Physics and Astronomy, University of Pittsburgh, 3941 O’Hara St, Pittsburgh, PA 15260, USA}

\author[0000-0002-5653-0786]{Paul Torrey}
\affiliation{Department of Astronomy, University of Virginia, 530 McCormick Road, Charlottesville, VA 22903, USA}

\begin{abstract}

We present a suite of high-resolution numerical simulations to study the evolution and survival of dust in hot galactic winds. We implement a novel dust framework in the Cholla hydrodynamics code and use wind tunnel simulations of cool, dusty clouds to understand how thermal sputtering affects the dust content of galactic winds. Our simulations illustrate how various regimes of cloud evolution impact dust survival, dependent on cloud size, wind properties, and dust grain size. We find that significant amounts of dust can survive in winds in all scenarios, even without shielding from the cool phase of outflows. We present an analytic framework that explains this result, along with an analysis of the impact of cloud evolution on the total fraction of dust survival. Using these results, we estimate that 60 percent of 0.1 micron dust that enters a starburst-driven wind could survive to populate both the hot and cool phases of the halo, based on a simulated distribution of cloud properties. We also investigate how these conclusions depend on grain size, exploring grains from 0.1 micron to 10 Angstrom. Under most circumstances, grains smaller than 0.01 micron cannot withstand hot-phase exposure, suggesting that the small grains observed in the CGM are either formed in situ due to the shattering of larger grains, or must be carried there in the cool phase of outflows. Finally, we show that the dust-to-gas ratio of clouds declines as a function of distance from the galaxy due to cloud-wind mixing and condensation. These results provide an explanation for the vast amounts of dust observed in the CGMs of galaxies and beyond.

\end{abstract}


\section{Introduction} \label{sec:introduction}

Dust and galaxy evolution are entwined. The presence of dust in the ISM can enhance star formation rates in galaxies, because dust serves as a cooling channel for gas \citep{Whitworth1998}, a catalyzing surface for the formation of molecular hydrogen \citep{Hollenbach1971}, and a shield from molecular photodissociation by ultraviolet starlight \citep{Draine1996}---all of which aid in galaxies' abilities to form stars. However, dust grains have also been observed to hold a significant portion of the ISM's metal content \citep{Dwek1998, Weingartner2001} and have the potential to accrete metals from the gas phase \citep{Dwek1998, Hirashita1999, Jenkins2009}, which could decrease the efficiency of gas cooling. Thus, an understanding of cosmic dust evolution in galaxies is critical to an overall theory of galaxy evolution.

A growing number of observations show that extragalactic dust is abundant in the Universe \citep{Zaritsky1994, Chelouche2007, Menard2010, McGee2010, Menard2012, Peeples2014, HodgesKluck2014, Peek2015, Smith2016}. For example, using reddening in observations of gravitationally lensed quasars in galaxy halos, \cite{Menard2010} found that extragalactic dust accounts for roughly half of the cosmic dust mass and can extend out to Mpc scales. However, the source of such vast amounts of extragalactic dust is unclear, given that dust is thought to form predominantly in the interstellar medium (ISM) through condensation in stellar winds and supernovae \citep{Dwek1980, Gehrz1989, Moseley1989}.

Mechanisms that may explain in situ extragalactic dust formation have been suggested \citep[e.g.][]{Draine1990}, but it is unclear how efficiently these mechanisms operate on a broad scale. Thus, various channels for transporting dust out of the ISM have been proposed to explain this phenomenon. These include radiation pressure-driven outflows \citep[e.g.][]{Davies1998, Murray2005}, tidal stripping events \citep[e.g.][]{Melendez2015, Yoon2021}, and hot supernova-driven outflows \citep[e.g.][]{Shopbell1998}. It remains uncertain, however, the extent to which each of these may contribute to the observed extragalactic cosmic dust mass.

It is also unclear how the morphological properties of dust grains may be altered through these processes. In the ISM, dust grains range in diameter from nanometers to microns \citep{Weingartner2001}. Their sizes can also change with time---dust can grow through coagulation \citep{Asano2013, Hirashita2023} and shrink due to shattering \citep{Jones1996, Asano2013}. It can also be destroyed altogether due to sputtering \citep{Draine1979}. The efficiency of these mechanisms in outflows is unknown. In particular, observations suggest that the abundance of small grains in the circumgalactic medium (CGM) is comparable to that of the ISM \citep{Wolfire1995}. At first glance, this is puzzling, since theoretical models predict that the virial temperatures of massive galaxies should be hot enough to destroy small grains \citep{Spitzer1956, Strickland2004, Anderson2010}. It is unclear whether the source of extragalactic dust directly supplies small grains to the CGM or if they are formed in situ through shattering \citep{Jones1996}.

Star formation feedback-driven outflows are a promising explanation for the observed ubiquity of dusty halos in the Universe. These outflows are often observed in rapidly star-forming galaxies, both at low and high redshift \citep[e.g.][and references therein]{Heckman1990, Veilleux2005, Veilleux2020}. Hot, galaxy-scale outflows are generated by supernova explosions \citep{Chevalier1985}, which can sweep up cool, dense clouds of gas that break out from the disk \citep[e.g.][]{MacLow1988, Sarkar2015, Tanner2016, Kim2017, Fielding2018, Schneider2018b, Nguyen2022}. These outflows are frequently observed in the form of blueshifted absorption line spectra, which predominantly probe outflowing cool ($\sim10^4~\textrm{K}$) gas moving at speeds higher than the escape velocity, thought to originate from the sites of supernovae \citep[e.g.][]{Martin1998, Rupke2005, Weiner2009, Rubin2010, Rubin2014}. Numerical hydrodynamics simulations have demonstrated that these clouds can be long-lived \citep[e.g.][]{Cooper2009, Gronke2018, Schneider2020, Abruzzo2022a, Schneider2024}, and thus may be a source of the significant amounts of cool gas observed in the CGM \citep{Werk2013}. Logically, clouds that are launched from dusty sites of star formation may also supply the CGM with dust.

Despite this, the ability of galactic outflows to transport dust to the CGM is debated. Although there is observational evidence that clouds in outflows can be dusty \citep{BlandHawthorn2003, Rupke2013, Triani2021, Katsioli2023}, the evolution of dust in outflows is not well understood. In particular, dust in the presence of the extremely hot gas that drives outflows may be susceptible to sputtering: the gradual degradation of dust grains as collisions with gas ions return bound atoms and molecules to the gas phase. This phenomenon is invoked to explain dust depletion in supernova remnants, which is thought to be caused in part by efficient sputtering of dust grains in shocked gas \citep[e.g.][]{Jones1994, Dopita2016}.

By volume, outflows are thought to consist mostly of hot ($\sim10^7~\textrm{K}$) ionized gas, as observed in the nearby starburst galaxy M82 \citep{Griffiths2000, Strickland2007, Lopez2020} and shown in simulations \citep{Kim2017, Schneider2020}. They are also turbulent, as discontinuities in density, temperature, and momentum at the boundary between cool clouds and the wind drive mixing between phases \citep{Fielding2020, Tan2021, Abruzzo2022b}. Dust grains are thought to be easily sputtered in these conditions due to the increased frequency of gas-grain collisions due to gas thermal motions \citep{Draine1979}. The same is true of turbulent gas since non-zero gas-grain relative velocities can arise \citep{Hirashita2009}.

The significance of sputtering's role in dust survival in outflows has not been thoroughly explored. Single-phase numerical simulations of the effect of dust-enhanced cooling on the thermal properties of hot galactic winds have shown that, in most cases, sputtering is a significant source of dust destruction on timescales shorter than the outflow dynamical time \citep{Ferrara2016, Scannapieco2017}. It has been proposed that efficient shielding of dust by the cool phase of outflows may enable long-term dust survival. However, this latter mechanism has only begun to be explored in simulations \citep{Farber2022, Chen2023, Otsuki2024}. 

In this work, we address this question by conducting high-resolution simulations of individual dusty clouds as they are accelerated by hot galactic winds, enabling direct measurements of the dust content of outflows as it evolves due to sputtering. In Section~\ref{sec:analytics}, we make an analytic argument for the survival of dust in outflows based on typical outflow properties and dust sputtering times. In Section~\ref{sec:simulations}, we describe the implementation of a numerical dust model in the Cholla code \citet{Schneider2015}, and describe our simulation setup. We then perform a parameter study to demonstrate how cloud size, cloud and wind properties, and dust grain size affect dust survival in winds, presented in Section~\ref{sec:results}. These results provide insights into the dust-to-gas ratio and grain size distribution of clouds in outflows and also allow us to put upper limits on the CGM dust mass transported in outflows, which we discuss in Section~\ref{sec:discussion}. We conclude in Section~\ref{sec:conclusions}.

\section{Analytics} \label{sec:analytics}
In this Section, we make a simple analytic argument for the survival of dust in galactic outflows. This model shows that the timescale for dust destruction is almost always much longer than the overall time of cloud-wind dynamical evolution. First, we describe the primary mechanism responsible for dust destruction in winds, thermal sputtering. Then, we examine the sputtering times in the cool, mixed, and hot phases of outflows. We show that, in general, outflow sputtering times are quite long---especially in the phases where most dust can be found. We will discuss several dust evolution scenarios based on cloud survivability.

\subsection{Sputtering} \label{subsec:destruction-mechanisms}

Assuming that dust is spherical and has a constant density, the mass of a dust grain is given by $m=4\pi a^3 \rho_\mathrm{g}/3$, where $m$ is the grain mass, $a$ is the grain radius, and $\rho_\mathrm{g}$ is the grain density. Using this, we can describe the rate of change in dust mass as

\begin{equation}
\Big|\frac{\dot{m}}{m}\Big|=3\frac{\dot{a}}{a}
\label{eq:sput-deriv}
\end{equation}

\noindent which leads to the general definition of the sputtering time. The sputtering time quantifies the time it takes for a dust grain to be destroyed by sputtering,

\begin{equation}
t_\mathrm{sp}=a\Big|\frac{\textrm{d}a}{\textrm{d}t}\Big|^{-1}.
\label{eq:sput-analytic}
\end{equation}

\noindent Given this, the rate of change in dust density due to sputtering can be written as

\begin{equation}
\frac{\textrm{d}\rho}{\textrm{d}t}=-\frac{\rho}{t_\mathrm{sp}/3}.
\label{eq:sput-model}
\end{equation}

How rapidly dust grains decay through sputtering depends on the sputtering time. \citet{Draine1979} provided an analytic expression for the sputtering rate, $\textrm{d}a/\textrm{d}t$, that accounts for the transfer of kinetic energy from the impinging particles and electric field effects that both play a part in sputtering. Using the assumption that the decay rate of a given solid does not depend on the type of incident particle \citep{Bohdansky1984}, material-independent sputtering yields have been measured, which depend only on the thermal properties of the impinging particles. This allowed for greatly simplified, semi-empirical sputtering models. The sputtering rate as a function of sputtering yield for astrophysically-relevant materials is given by

\begin{equation}
\frac{1}{n_\mathrm{H}}\frac{\textrm{d}a}{\textrm{d}t}=\frac{m_\mathrm{sp}}{2\rho_0}\sum_i A_i \langle Y_i v\rangle 
\label{eq:sput-empirical}
\end{equation}

\noindent (thermal sputtering, \citealt{Tielens1994}). Here, $\dot{a}$ depends on the abundance of ion $i$, $A_i$, and its empirically measured sputtering yield averaged over a Maxwellian distribution, $\langle Y_i v\rangle$. $m_\mathrm{sp}$ and $\rho_0$ are the average mass of the sputtered atoms and the specific density of the grain material, respectively. \citet{Tielens1994} provides a similar treatment for non-thermal (or inertial) sputtering, but since gas-grain recoupling in outflows happens quickly compared to non-thermal sputtering (as discussed in Appendix~\ref{app:nonth}), we neglect it in this work.

Using these sputtering yields, the thermal sputtering rate can be written as 

\begin{equation}
\frac{\textrm{d}a}{\textrm{d}t}=-\tilde{h}\Bigg(\frac{\rho}{m_\mathrm{p}}\Bigg)\Bigg[\Bigg(\frac{T_\mathrm{sp}}{T}\Bigg)^\omega+1\Bigg]^{-1}.
\label{eq:sput-fit}
\end{equation}

\noindent \citep{Tsai1995}. For graphite, amorphous carbon, and silicate grains, $\bar{h}=3.2\times10^{-18}~\textrm{cm}^4\,\textrm{s}^{-1}$, $\omega=2.5$, and $T_\mathrm{sp}=2\times10^6~\textrm{K}$. Finally, the thermal sputtering time as a function of grain radius, $a$, gas density, $\rho$, and gas temperature, $T$, can be written as

\begin{equation}
t_\mathrm{sp}\approx0.17~\textrm{Gyr}\Big(\frac{a}{0.1~\mu\text{m}}\Big)\Big(\frac{10^{-27}~\textrm{g}\,\textrm{cm}^{-3}}{\rho}\Big) \Big[\Big(\frac{10^{6.3}~\textrm{K}}{T}\Big)^\omega+1\Big]
\label{eq:sput-timescale}
\end{equation}

\noindent \citep{McKinnon2017}. We implement Equation \ref{eq:sput-model} as our sputtering model with Equation \ref{eq:sput-timescale} as our sputtering time. Equation \ref{eq:sput-timescale} shows that sputtering is most efficient in high-density, high-temperature gas since $t_\mathrm{sp}$ decreases with increased density and temperature. It depends most strongly on temperature and is roughly constant above $T_\mathrm{sp}$ (which is set by the constant $\omega$). Large dust grains can withstand sputtering for longer, so $t_\mathrm{sp}$ scales linearly with grain radius. In Figure \ref{fig:t_sp}, we plot contours showing the sputtering time as a function of density and temperature for $a=0.1~{\mu\textrm{m}}$ grains. This Figure illustrates that sputtering times for grains this size are short ($\lesssim1~\textrm{Myr}$) only in gas that is both very hot \emph{and} dense.

\subsection{Sputtering in multiphase outflows} \label{subsec:multiphase}

Gas in galactic outflows is inherently multiphase. Although there is a continuous range of densities and temperatures, gas is commonly divided into several distinct phases based on its cooling efficiency at different temperatures. The hot phase, which we refer to as the ``wind," is the fast, volume-filling phase that drives the expansion of the outflow. It is created by the thermalization of supernovae ejecta in the ISM and is characterized by high temperatures and relatively diffuse densities \citep[e.g.][]{Strickland2007}. Embedded in the hot wind are cooler, denser clouds of $\sim10^{4}\,\mathrm{K}$ gas, which are launched from the ISM by interactions with supernova bubbles near star-forming regions \citep[e.g.][]{Westmoquette2009}. These clouds make up the cool ionized phase of outflows and likely dominate the mass that is driven out of galaxies \citep{Kim2020}. Finally, the interaction between the hot wind and cool clouds creates an intermediate temperature and density phase with relatively short cooling times. This phase forms in the outflow as the wind sweeps along the boundaries of clouds, seeding hydrodynamic instabilities that initiate cloud-wind mixing \citep{Fielding2020}. The phases of outflows are characterized by their distinct thermodynamic properties, which by definition lead to different sputtering times in each phase. In this Section, we examine how dust evolves in each phase.

\begin{figure}
\centering
\includegraphics[width=0.43\textwidth]{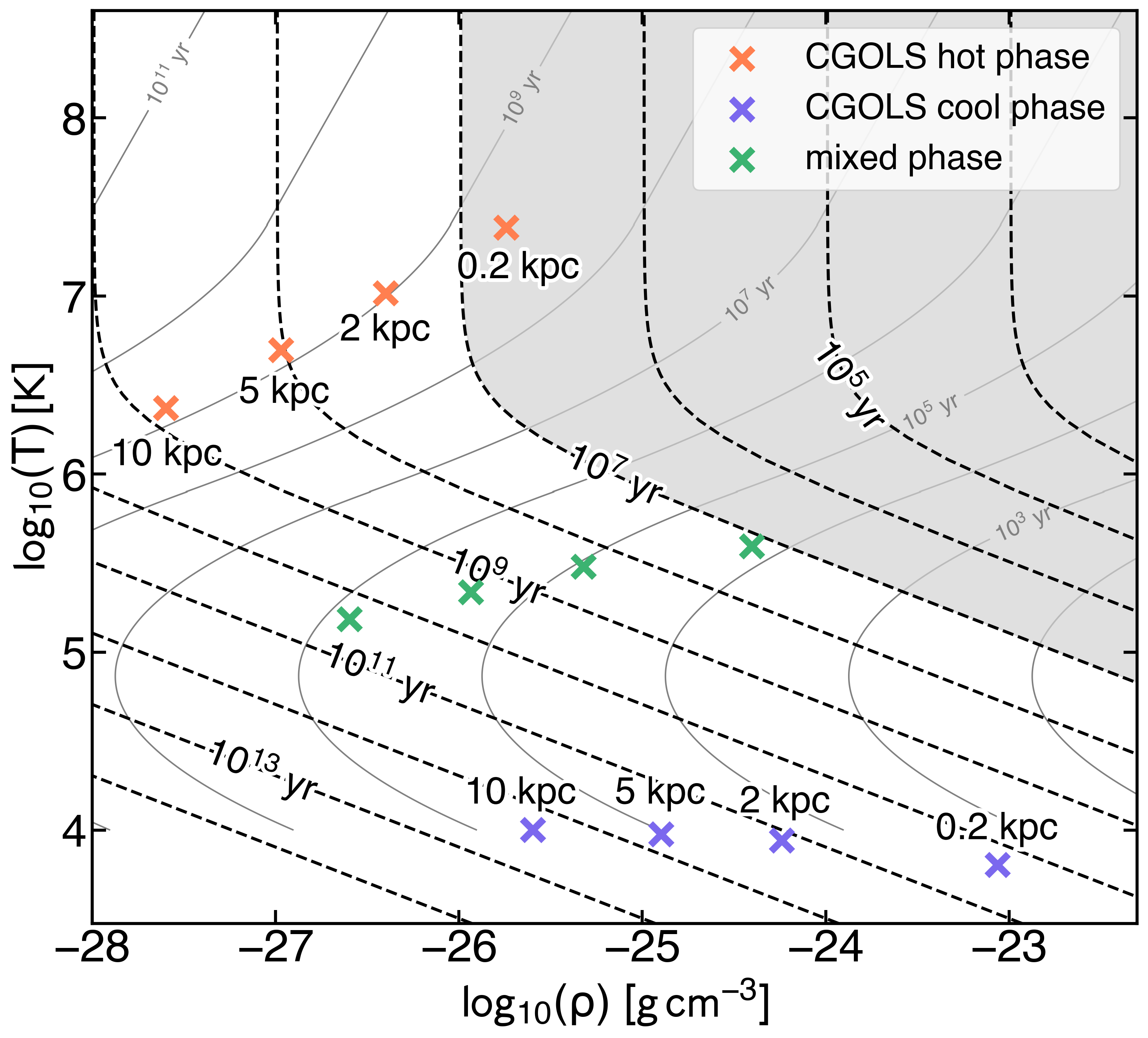}
\caption{Contours (dashed black lines) showing sputtering times in years for $a=0.1~\mu\textrm{m}$ dust grains over a range of densities and temperatures. The labeled points mark the median values of the hot ($T_\mathrm{w}>5\times10^5~\textrm{K}$), cool ($T_\mathrm{cl}<2\times10^4~\textrm{K}$), and mixed ($T_\mathrm{mix}=(T_\mathrm{cl}T_\mathrm{w})^{1/2}$) phases of the CGOLS profiles at different distances from the outflow's base. The gray shaded region highlights where sputtering times are shorter than the time it takes for a $10^3~{\textrm{km}\,\textrm{s}^{-1}}$ wind to travel $10~\textrm{kpc}$. For all phases, sputtering times quickly lengthen with distance from the galaxy due to the rapid drop in temperature and density. In particular, even the hot phase sputtering times are quite long, surpassing $10~\textrm{Myr}$ by a distance of $1~\textrm{kpc}$ from the base of the wind and $100~\textrm{Myr}$ by a distance of $5~\textrm{kpc}$. The solid gray contours show the CIE cooling time of gas in years.}
\label{fig:t_sp}
\end{figure}

Throughout this work, we refer to the Cholla Galactic OutfLow Simulations (CGOLS) project to characterize the properties of hot, cool, and mixed phases of outflows. CGOLS is a suite of simulations of rapidly star-forming galaxies that drive multi-phase outflows \citep{Schneider2018a, Schneider2020, Schneider2024}. CGOLS provides a detailed understanding of the phase structure of outflows, including profiles of the density, temperature, velocities, etc. of hot ($T>5\times10^5~\textrm{K}$) and cool ($T<2\times10^4~\textrm{K}$) phases as a function of distance from the galaxy, which we use as a reference. Mean values from these profiles are shown as colored points in Figure \ref{fig:t_sp}.

\subsubsection{Cool phase} \label{subsubsec:coolphase}

Clouds in CGOLS typically have a temperature of $T_\mathrm{cl}<2\times10^4~\textrm{K}$ and number densities that range between $10^2~\textrm{cm}^{-3}$ near the base of the outflow and $10^{-2}~\textrm{cm}^{-3}$ further away from the galaxy. In general, cool phase sputtering times are long. This is indicated in Figure \ref{fig:t_sp}, where the CGOLS radial profiles for the cool phase are shown as a function of density and temperature, along with contours indicating the corresponding sputtering times for a $0.1~{\mu\textrm{m}}$ radius grain. The shortest cool phase sputtering time for a dust grain of this size is roughly $10~\textrm{Gyr}$, approximately a Hubble time. As such, essentially no sputtering of $a=0.1~{\mu\textrm{m}}$ dust will occur in cool-phase gas. Although $t_\mathrm{sp}$ decreases linearly with grain size, smaller grains may also be able to survive in clouds, since their sputtering times are still long compared to outflow dynamical times and cloud density decreases as clouds are carried out.

\subsubsection{Hot phase} \label{subsubsec:hotphase}

The hot phase is characterized by high temperatures ($T_\mathrm{w}\gtrsim5\times10^5~\textrm{K}$) and long cooling times, with typical densities of order $n_\mathrm{w}\sim10^{-2}-10^{-4}~\textrm{cm}^{-3}$. Wind densities and temperatures are both highest at the base of the outflow, nearest to the supernovae that drive them. There, $t_{\mathrm{sp,w}}\sim6~\textrm{Myr}$ for $a=0.1~{\mu\textrm{m}}$ grains---significantly shorter than the cool phase sputtering times. If dust spends longer than $t_{\mathrm{sp,w}}$ in the hot phase, substantial amounts could be destroyed. To quantify dust's ability to survive in the hot phase, we can define the sputtering distance, $r_\mathrm{sp}=t_\mathrm{sp}\times v_\mathrm{w}$, to be the distance dust would travel moving at the wind speed before being significantly sputtered. For a sputtering timescale of $6~\textrm{Myr}$, a typical wind speed of $v_\mathrm{w}=10^3~{\textrm{km}\,\textrm{s}^{-1}}$ would result in a sputtering distance of $6~\textrm{kpc}$. This distance is small compared to the scale of galactic outflows, which can span tens of kpc. However, the sputtering time of the hot phase does not remain constant throughout outflows. Figure \ref{fig:t_sp} shows, for example, the CGOLS hot phase sputtering time at various points in the wind. The CGOLS wind density and temperature profiles decrease rapidly with distance from the galaxy, causing the sputtering time to double by a distance of $1~\textrm{kpc}$. At a distance of $5~\textrm{kpc}$ from the outflow's base, the hot phase sputtering time increases by over an order of magnitude. With this consideration, it is plausible that some  $a=0.1~{\mu\textrm{m}}$ dust could survive moving in a rapid wind fully exposed to the hot phase. In the case of smaller grains, the sputtering distance is quite short since the sputtering time depends linearly on grain size. For $a=0.01~{\mu\textrm{m}}$ grains, the sputtering distance is $0.6~\textrm{kpc}$, so grains of this size and smaller would likely be completely sputtered before reaching regions of the wind with longer sputtering times.

\subsubsection{Mixed phase} \label{subsubsec:mixedphase}

Despite high temperatures, the relatively diffuse densities of the hot phase result in long sputtering times for large grains. In principle, the intermediate temperatures \emph{and} densities of the mixed phase could be more amenable to sputtering. Additionally, cooler gas tends to move less quickly, which increases the exposure time for dust in this phase. To define the thermodynamic properties of the mixed phase, we use the geometric means of the hot and cool phase temperature and density: $T_\mathrm{mix}=(T_\mathrm{cl}T_\mathrm{w})^{1/2}$ and $n_\mathrm{mix}=(n_\mathrm{cl}n_\mathrm{w})^{1/2}$, respectively. 
The shortest sputtering time for this phase is therefore also at the base of the outflow; with $n_\mathrm{mix}\sim0.4~\textrm{cm}^{-3}$ and $T_\mathrm{mix}\sim10^5~\textrm{K}$ we get a sputtering time of $15~\textrm{Myr}$. However, this is much longer than the mixed phase cooling time, which governs how long mixed phase gas exists. The cooling timescale is defined as

\begin{equation}
    t_\mathrm{cool}=\frac{3 n k T}{2n_\mathrm{H}^2\Lambda(T)},
\end{equation}

\noindent where $n$ is number density, $n_\mathrm{H}$ is the Hydrogen number density, $\Lambda(T)$ is the cooling function in $\textrm{erg}\,\textrm{s}^{-1}\,\textrm{cm}^{3}$, and $k$ is the Boltzmann constant. Mixed-phase gas in outflows is typically around $\sim10^5~\textrm{K}$, which is close to the maximum $\Lambda$ for gas in collisionally ionized equilibrium (see contours in Figure \ref{fig:t_sp}). This results in cooling times of order $\sim10~\textrm{kyr}$ for mixed-phase gas in outflows \citep{Gronke2018}. As a result, we do not expect substantial sputtering of dust in the mixed phase of outflows for $0.1~{\mu\textrm{m}}$ radius grains. That said, the sputtering decreases with grain size, so smaller grains ($a\sim10~\textup{\AA}$) can be susceptible to sputtering in the mixed phase.

\subsection{Cloud crushing} \label{subsec:cloudcrushing}

In the preceding section, we showed that some dust may be sputtered in the hot phase of outflows, particularly small grains. However, clouds of cool gas should be very efficient at preventing even small dust grains from sputtering. Given this, we now turn to a discussion of cloud evolution in outflows. In particular, because the sputtering times of dust in clouds are long, if clouds can survive intact as they are carried to the CGM by the wind, then the vast majority of dust should survive as well.

Much work has been done to quantify the question of cloud survival in winds \citep{Klein1994, Scannapieco2015, Schneider2017, Gronke2018, Li2020, Sparre2020, Farber2022, Abruzzo2023}. In particular, many studies have used the framework of the cloud-crushing problem, which considers the evolution of cool, dense clouds of gas as they are accelerated from rest by hot, relatively diffuse winds. The adiabatic cloud-crushing time, $t_\mathrm{cc}$, has been defined to quantify this evolution \citep{Klein1994}. This is the time it takes for the initial cloud-wind shock to propagate through the cloud:

\begin{equation}
t_\mathrm{cc}=\frac{\chi^{1/2} r_\mathrm{cl}}{v_\mathrm{w}},
\label{eq:tcc_adiabatic}
\end{equation}

\noindent where $\chi$ is the density contrast between the cloud and the wind, $v_\mathrm{w}$ is the wind velocity, and $r_\mathrm{cl}$ is the initial radius of the cloud. In the adiabatic limit (i.e. when radiative cooling is subdominant), cloud evolution is dominated by the growth of Kelvin-Helmholtz and Rayleigh-Taylor instabilities that eventually shred the cloud apart. The adiabatic cloud-crushing time is comparable to the timescale for this growth and thus can be used to quantify the general sequence of cloud evolution.

\begin{table*}[t]
  {\centering
  \begin{tabular}{c c c c c c c c c c }
  \toprule
    Resolution & Dimensions~($r_\mathrm{cl}$) & $r_\mathrm{cl}~(\textrm{pc})$ & $a~(\mu\textrm{m})$ & $v_\mathrm{w}~(\textrm{km}/\textrm{s})$ & $T_\mathrm{w}~(\textrm{K})$ & $\chi$ & $m_\mathrm{d,sp}/m_\mathrm{d,i}$ & $t/t_\mathrm{sp,w}$ & $t_\mathrm{cool,minmix}/t_\mathrm{shear}$ \\
  \midrule

    $r_\mathrm{cl}$/64 & $100\times20\times20$ & 5 & 0.1 & $1\times10^3$ & $3\times10^7$ & $10^3$ & 0.24 & 0.29 & 33 \\

    $r_\mathrm{cl}$/16 & \textquotesingle\textquotesingle & \textquotesingle\textquotesingle & 0.01 & \textquotesingle\textquotesingle & \textquotesingle\textquotesingle & \textquotesingle\textquotesingle & 0.88 & 3.1 & \textquotesingle\textquotesingle \\

    \hline

    $r_\mathrm{cl}$/64 & $128\times16\times16$ & 100 & 0.1 & $1\times10^3$ & $3\times10^7$ & $10^3$ & 0.28 & 5.3 & 1.7 \\

    $r_\mathrm{cl}$/16 & $64\times16\times16$ & \textquotesingle\textquotesingle & 0.01 & \textquotesingle\textquotesingle & \textquotesingle\textquotesingle & \textquotesingle\textquotesingle & 0.73 & 49 & \textquotesingle\textquotesingle \\

    \hline
    
    $r_\mathrm{cl}$/64 & $128\times16\times16$ & 100 & 0.1 & $5\times10^2$ & $3\times10^6$ & $10^3$ & 0.01 & 5.6 & 0.22 \\

    $r_\mathrm{cl}$/16 & $64\times16\times16$ & \textquotesingle\textquotesingle & 0.01 & \textquotesingle\textquotesingle & \textquotesingle\textquotesingle & \textquotesingle\textquotesingle &  0.08 & 41 & \textquotesingle\textquotesingle \\

    $r_\mathrm{cl}$/16 & \textquotesingle\textquotesingle & \textquotesingle\textquotesingle & 0.001 & \textquotesingle\textquotesingle & \textquotesingle\textquotesingle & \textquotesingle\textquotesingle & 0.53 & 580 & \textquotesingle\textquotesingle \\

  \bottomrule
  \end{tabular}
  \caption{Parameters for simulations discussed in Section \ref{sec:results}. Resolution is the number of cells per cloud radius, dimensions are the x, y, and z lengths of the simulation domain (in units of the initial cloud radius), $r_\mathrm{cl}$ is the cloud radius, $a$ is the dust grain size, $v_\mathrm{w}$ is the wind velocity, $T_\mathrm{w}$ is the wind temperature, $\chi$ is the cloud-wind density contrast, $m_\mathrm{d,sp}/m_\mathrm{d,i}$ is the fraction of total sputtered dust compared to the initial total dust mass, $t/t_\mathrm{sp}$ is the simulation duration compared to the sputtering time, and $t_\mathrm{cool,minmix}/t_\mathrm{shear}$ is the mixed-phase cooling time compared to the cloud's shear time. The horizontal lines divide the table into three regimes: destroyed (top), disrupted (middle), and survived cloud (bottom). Our simulation gallery can be found at \url{https://vimeo.com/showcase/11013652}.}
  \label{tab:sims}}
\end{table*}

Clouds in the adiabatic case generally survive for $\sim5~t_\mathrm{cc}$ before they are completely mixed into the wind \citep[][etc.]{Klein1994, Schneider2015}. Equation \ref{eq:tcc_adiabatic} shows that small clouds are disrupted more quickly than large ones and, generally, that clouds are disrupted more quickly by fast winds. For example, in a $10^3~{\textrm{km}\,\textrm{s}^{-1}}$ wind, a typical large cloud ($r_\mathrm{cl}\sim100~\textrm{pc}$ and $\chi\sim10^3$) will get disrupted on a timescale of $\sim3~\textrm{Myr}$, while a small cloud ($r_\mathrm{cl}\sim5~\textrm{pc}$ and $\chi\sim10^3$) will be disrupted in approximately $155~\textrm{kyr}$. Were the adiabatic case the final word, both of these clouds would be destroyed in a galactic outflow, since $t_\mathrm{cc}$ is much shorter than the outflow dynamical time.

In the context of outflows, however, adiabaticity is a poor approximation since gas can cool on much shorter timescales than the outflow dynamical time. Recent work has shown that rapid cooling of the mixed phase significantly alters the evolution of clouds by extending their lifetimes or preventing their destruction, entirely \citep[e.g.][]{Marinacci2010, Armillotta2016, Gritton2017, Gronke2018, Gronke2020a, Li2020, Sparre2020}. For the small and large clouds discussed above with a temperature of $T_\mathrm{cl}=3\times10^4~\textrm{K}$ in a wind of $T_\mathrm{w}=3\times10^7~\textrm{K}$ and $n_\mathrm{w}=10^{-2}~\textrm{cm}^{-3}$, the mixed-phase cooling time is $t_\mathrm{cool,mix}= 209~\textrm{kyr}$. The small cloud's $t_\mathrm{cc}$ is comparable to the mixed-phase cooling time, so cooling will extend the cloud's lifetime to some degree. The large cloud's $t_\mathrm{cc}$ is much longer than $t_\mathrm{cool,mix}$, so cooling will have a significant effect on its evolution.

Cloud survival is tied to the mixed gas phase, which is composed of a mixture of disrupted cloud material and material from the hot phase. Clouds can absorb rapidly cooled gas from this mixed phase, which, in some cases, can facilitate cloud survival through a process called turbulent radiative mixing layer (TRML) entrainment. A cloud survives when it absorbs more material than it contributes to the mixed phase. In other words, survival requires the absorption of material originating from the hot phase. The mixed material can only be absorbed as it advects the length of the cloud---if it advects beyond this, it will be lost to mixing with the wind. Consequently, the comparison between the cooling timescale of mixed gas and the advection timescale of mixed gas originating from the wind dictates cloud survival.

In more detail, this advection timescale is linked to the shear timescale,

\begin{equation}
t_\mathrm{shear}=r_\mathrm{cl}/v_\mathrm{w},
\label{eq:crit}
\end{equation}

\noindent which quantifies the time it takes for a parcel of gas moving at the wind speed, $v_\mathrm{w}$, to travel half the initial cloud length, $r_\mathrm{cl}$.
In reality, the advection timescale is somewhat longer because the cloud develops a tail and mixing reduces the speed. \citet{Abruzzo2023} defined the cloud survival criterion,

\begin{equation}
\label{eq:survival-criterion}
t_\mathrm{cool,minmix}\lesssim7~t_\mathrm{shear}.
\end{equation}

\noindent In this criterion, $t_\mathrm{cool,minmix}$ is a quantity similar in spirit to the cooling time of the mixing layer. It is the cooling time at $T=\sqrt{T_\mathrm{min} T_\mathrm{w}}$, where $T_\mathrm{min}$ is the temperature between $T_{\rm cl}$ and $T_{\rm w}$ where the cooling time is minimized ($\sim3\times 10^4~\textrm{K}$ in our simulations). Throughout this paper, we use this criterion to quantify cloud survival. We define three regimes of cloud evolution that demonstrate distinct characteristics of dust survival: cloud destruction ($t_\mathrm{cool,minmix}>t_\mathrm{shear}$), disruption ($t_\mathrm{cool,minmix}\sim t_\mathrm{shear}$) and survival ($t_\mathrm{cool,minmix}<t_\mathrm{shear}$). In Section~\ref{sec:results}, we present simulations of dusty clouds in each of these regimes and discuss how the survival of dust is affected by cloud dynamics.

\begin{figure*}
\centering
\includegraphics[width=\textwidth]{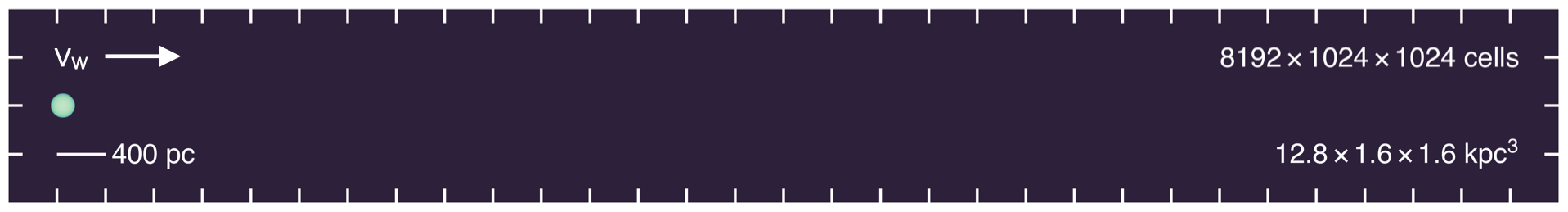}
\caption{Simulation domain for our large cloud simulations. Here, we simulate a $100~\textrm{pc}$ radius cloud in a box that is $128~r_\mathrm{cl}$ long at a fixed resolution of $r_\mathrm{cl}/64$ cells ($\textrm{d}x=1.5625~\textrm{pc}$). Box sizes for our other setups can be found in Table~\ref{tab:sims}.
\label{fig:initial_conditions}}
\end{figure*}

\section{Methods} \label{sec:simulations}

To test the survival of dust in a range of different outflow conditions, we ran a suite of hydrodynamic simulations consisting of three-dimensional boxes containing a cool, dense cloud of gas and dust embedded in a hot, diffuse background wind. All simulations were run with the GPU-based hydrodynamics code, Cholla \citep{Schneider2015}, using piecewise-parabolic interface reconstruction \citep{Fryxell2000}, the HLLC Riemann solver (\citealt{Toro1994, Batten1997}), and the second-order unsplit Van Leer integrator \citep{Stone2009}. The cooling function, $\Lambda(T)$, is a piece-wise parabolic fit to a collisional ionization equilibrium curve for solar metallicity gas, as defined in \cite{Schneider2018a}. Below $10^4~\textrm{K}$ the cooling rate is zero. We assume a mean molecular weight of 0.6 when converting between mass and number densities. We track the evolution of dust by computing the change in dust density of each cell using Equation \ref{eq:sput-model}. This model uses Cholla's passive scalar framework to advect dust along with gas as it evolves hydrodynamically. Thus, the model implicitly assumes that gas and dust are fully dynamically coupled.

\subsection{Simulation setup} \label{subsec:sim-setup}

We simulate the evolution of a cool, dense cloud of gas and dust in a long box containing a volume-filling hot, diffuse wind, commonly referred to as a ``wind tunnel" simulation (shown in Figure~\ref{fig:initial_conditions}). To emulate the conditions of a hot, supernova-driven outflow, all non-cloud regions of the simulation are initialized with a constant, positive x-velocity ($v_\mathrm{w}$) with a constant initial density and temperature. At the $-x$ boundary of the simulation volume, we apply a constant inflow boundary with the same initial conditions. All other boundaries of the volume use outflow boundary conditions. Initially, the wind is dust-free. Positioned adjacent to the $-x$ boundary and centered in the yz-plane is a cool, dense cloud of gas and dust, initialized with a 0.01 dust-to-gas ratio. The cloud is spherical, with a constant density and zero initial velocity. The cloud initial temperature is set such that it is in thermal pressure equilibrium with the wind.

Table \ref{tab:sims} shows the range of parameters for our fiducial simulations. Our fiducial cloud and wind parameters were chosen based on the cool and hot phase profiles from the CGOLS simulations \citep{Schneider2020}, but we vary them in some simulations to test a wider range of possible parameter space. The full range of parameters we explore is shown in Appendix~\ref{app:additional_sims}, Table~\ref{tab:sims_app}. Our CGOLS-like, starburst wind has a density of $n_\mathrm{w}=10^{-2}~\textrm{cm}^{-3}$, temperature of $T_\mathrm{w}=3\times10^7~\textrm{K}$ and a wind speed of $v_\mathrm{w}=10^3~{\textrm{km}\,\textrm{s}^{-1}}$. We also simulate a slower, cooler wind at a density of $n_\mathrm{w}=10^{-2}~\textrm{cm}^{-3}$, temperature of $T_\mathrm{w}=3\times10^6~\textrm{K}$, and speed $v_\mathrm{w}=500~{\textrm{km}\,\textrm{s}^{-1}}$. Despite the difference in temperature, because $t_\mathrm{sp}$ is roughly constant for temperatures above $T_\mathrm{sp}=2\times10^6~\textrm{K}$, the sputtering time the starburst and slow wind models are nearly the same: $t_\mathrm{sp}=10~\textrm{Myr}$ and $t_\mathrm{sp}=14~\textrm{Myr}$, respectively.

Within these two winds, we simulate clouds of various sizes and density contrasts. Our small cloud has a radius of $5~\textrm{pc}$, the minimum cloud size that is numerically resolved in the CGOLS simulations, and comparable to the minimum cloud size that is resolved in optical observations of M82 \citep{Shopbell1998}. Our large cloud has a radius of $100~\textrm{pc}$, chosen to coincide with the upper end of the CGOLS cloud size distribution (Warren et al. 2024, \emph{in prep}). Our fiducial simulations all contain $\chi=10^3$ clouds, but we show results from lower density contrast simulations in Table~\ref{tab:sims_app}.

We take the small cloud in the starburst wind, the large cloud in the starburst wind, and the large cloud in the slow wind to be our fiducial destroyed, survived, and disrupted simulations (all with density contrasts of $\chi=10^3$). We repeat each simulation for two different grain radii, $0.1~{\mu\textrm{m}}$ and $0.01~{\mu\textrm{m}}$, and shown an additional simulation with $10~\textup{\AA}$ grains for the survived cloud. Our fiducial large cloud simulations use box dimensions of $12.8\times1.6\times1.6~\textrm{kpc}^3$, which are the box dimensions shown in Figure~\ref{fig:initial_conditions}\footnote{We also repeated the survived cloud simulation with a cloud tracking method and found that the results for dust destruction were consistent with our untracked simulations. See Appendix~\ref{app:cloud_tracking} for details.}. Our fiducial destroyed cloud simulation uses a box of $0.5\times0.1\times0.1~\textrm{kpc}^3$. For all of the cloud regimes, we run our $a=0.1~{\mu\textrm{m}}$ simulations at a resolution of $r_\mathrm{cl}/64$, but find that the results do not depend strongly on resolution (see Appendix~\ref{app:convergence}), so we use a resolution of $r_\mathrm{cl}/16$ for the small grain simulations. We present the results of these simulations in Section~\ref{sec:results}.

\begin{figure*}
\begin{interactive}{animation}{movies/dest_movie.mp4}
\centering
\includegraphics[width=0.97\textwidth]{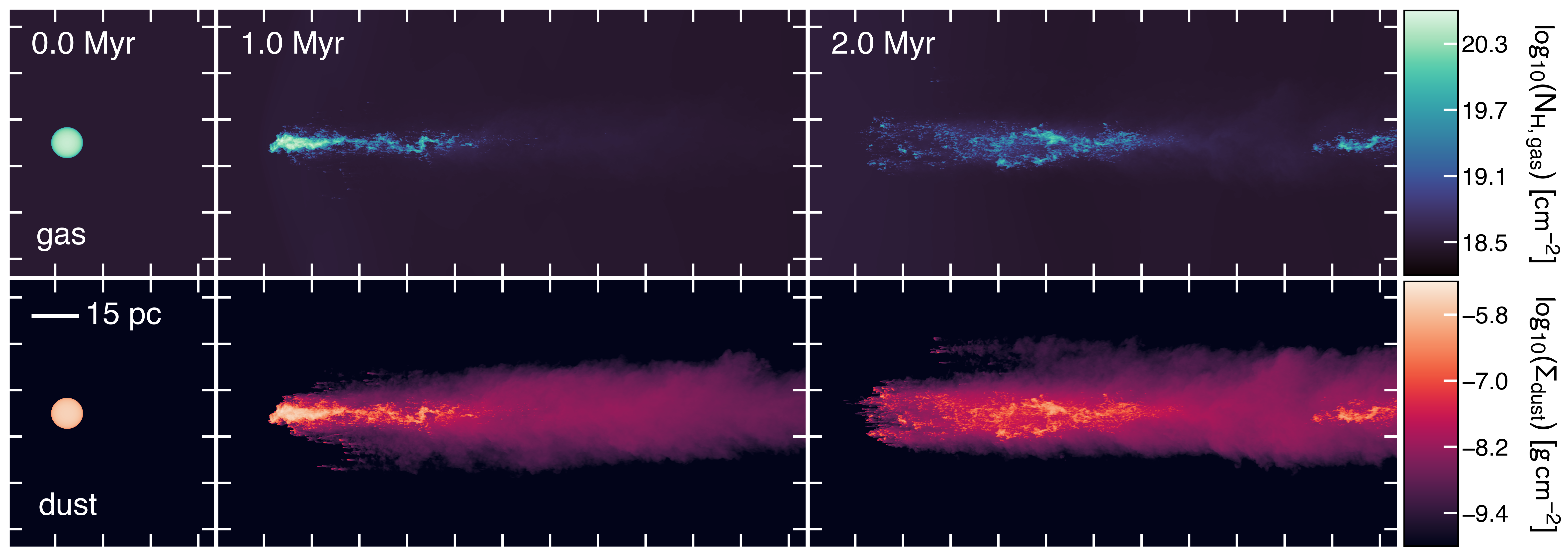}
\end{interactive}
\caption{Snapshots of the evolution of a destroyed cloud with $a=0.1~{\mu\textrm{m}}$ grains. The top panel shows gas column density and the bottom panel shows dust surface density. This simulation shows the evolution of a small cloud in our starburst wind model, such that $t_\mathrm{cool,minmix}>t_\mathrm{shear}$. As the cloud is disrupted by the wind, it fragments into small pieces that are subsequently mixed and heated into the wind. Cloud material travels nearly $500~\textrm{pc}$ over $2.9~\textrm{Myr}$ ($0.29~t_\mathrm{sp,w}$) by the time it leaves the box, by which point it is almost completely mixed into the hot phase. $24\%$ of the initial dust mass is sputtered. An animated version of this Figure can be found at \url{https://vimeo.com/showcase/11013652}.
\label{fig:destroyed-snapshots}}
\end{figure*}

\subsection{Dust mass analysis} \label{subsec:dust_mass_analysis}

For each simulation, we calculate the total fraction of dust that is sputtered, $m_\mathrm{d,sp}/m_\mathrm{d,i}$, which is shown in Table~\ref{tab:sims}. These masses are computed in situ from the sputtering rates in the simulation. To understand the degree to which hot phase sputtering affects the overall dust destruction, we separately track the total fraction of dust sputtered in $T<10^6~\textrm{K}$ and $T\geq10^6~\textrm{K}$ gas. We also estimate the total cloud and dust mass that have left the simulation volume in post-processing by calculating the mass outflow rate for each snapshot,

\begin{equation}
    \dot{m}=\rho v_\perp A,
\label{eq:outflow_rate}
\end{equation}

\noindent where $\dot{m}$ is the cloud gas (or dust) outflow rate, $\rho$ is the density of cloud gas (or dust) in a boundary cell, $v_\perp$ is the velocity in the direction flowing out of the volume, and $A$ is the area of the cell. Cloud gas is defined as $\rho>\rho_\mathrm{cl,i}/30$. In simulations where we initialize our cloud at $3\times10^3~\textrm{K}$, because there is no cooling to balance the effects of mixing at $T<10^4~\textrm{K}$, the temperature (density) of the colder phase gas will increase (decrease) by a factor of at least $10/3$. The amount of mass lost between each snapshot is calculated by multiplying $\dot{m}$ by the output timestep.

\begin{figure}
\centering
\includegraphics[width=0.42\textwidth]{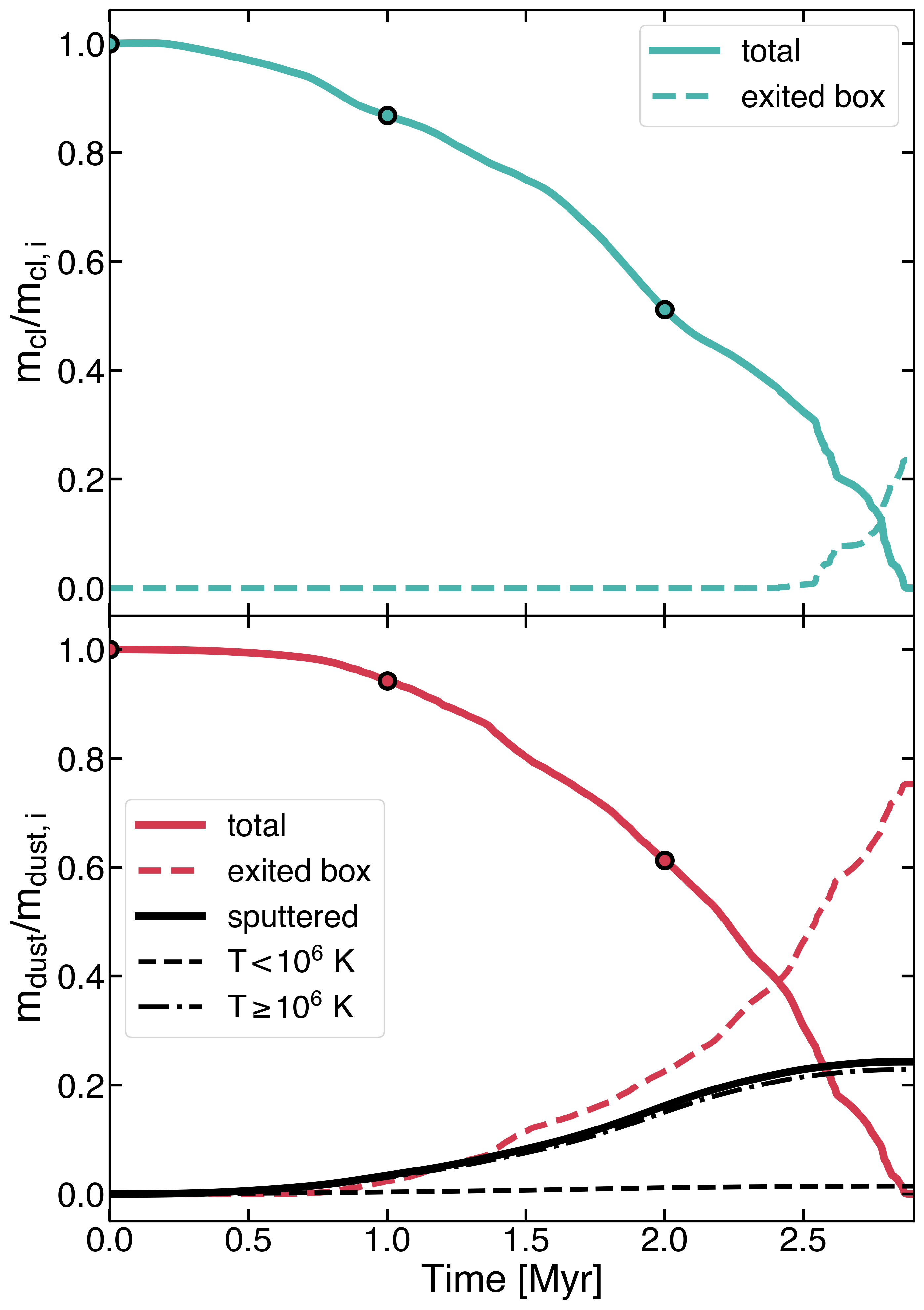}
\caption{\emph{Top:} Cloud mass evolution for a destroyed cloud simulation. The solid blue line shows the total amount of cloud gas in the volume relative to the initial cloud mass ($78~\textrm{M}_\odot$). The dashed blue line is an estimate of the total cloud mass that has been carried out of the volume (described in Section~\ref{subsec:dust_mass_analysis}).
\emph{Bottom:} Dust mass evolution for $a=0.1~{\mu\textrm{m}}$ grains in a destroyed cloud simulation. The solid red line is the total amount of dust in the simulation volume relative to the initial dust mass ($0.78~\textrm{M}_\odot$). The red dashed line is an estimate of the dust mass that has been carried out of the simulation. The solid black line is the total amount of dust sputtered. The black dashed and dot-dashed lines show the total amount of dust sputtered in $T<10^6~\textrm{K}$ and $T\geq10^6~\textrm{K}$, respectively. Overall, after $2.9~\textrm{Myr}$ ($0.29~t_\mathrm{sp,w}$), the cloud has been almost completely mixed into the wind and $24\%$ of the initial dust mass is sputtered, while the rest has been carried out of the simulation volume.
\label{fig:mass_dest}}
\end{figure}

\begin{figure}
\centering
\includegraphics[width=0.42\textwidth]{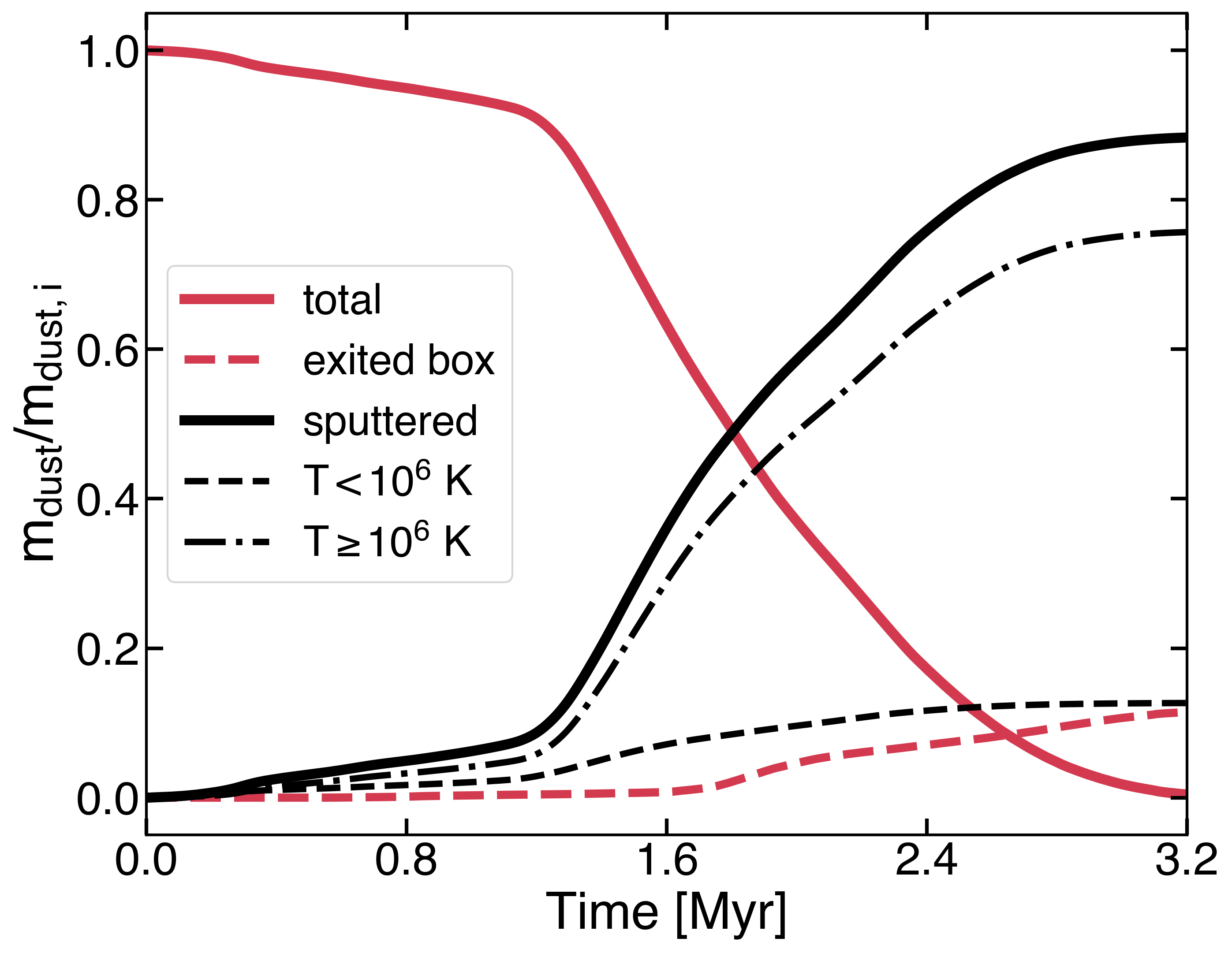}
\caption{Dust mass evolution for the same simulation setup as described in Figure~\ref{fig:mass_dest}, but with $a=0.01~{\mu\textrm{m}}$ dust grains. After roughly $3~\textrm{Myr}$, $12\%$ of the simulation's original dust content has exited the box, while $76\%$ of the dust is destroyed after mixing into $T\geq10^6~\textrm{K}$ gas, and $12\%$ of the dust is destroyed in cooler gas. The wind sputtering time for $a=0.01~{\mu\textrm{m}}$ grains is $1~\textrm{Myr}$.
\label{fig:mass_dest_small}}
\end{figure}

\section{Results} \label{sec:results}

In this Section, we present the results of our parameter study. Below, we highlight a selection of these simulations to illustrate how the extreme cases of cloud evolution, dust grain size, and efficient accretion of mixed-phase gas all play a role in overall dust survival. Simulations that are discussed in this section are shown in Table~\ref{tab:sims}.

\subsection{Destroyed clouds} \label{subsec:cloud-destruction}

To study dust survival in the case of complete cloud destruction, we simulate a small cloud in a hot, rapid wind, such that $t_\mathrm{shear}\ll t_\mathrm{cool,minmix}$. This scenario is meant to represent the outflow environment \emph{least} likely to result in dust survival. The cloud has a radius of $r_\mathrm{cl}=5~\textrm{pc}$ with a density contrast of $\chi=10^3$. We use the starburst wind described in Section~\ref{subsec:sim-setup}, which has a sputtering time of $10~\textrm{Myr}$ for $a=0.1~{\mu\textrm{m}}$ grains.

\begin{figure*}
\begin{interactive}{animation}{movies/surv_movie.mp4}
\centering
\includegraphics[width=0.97\textwidth]{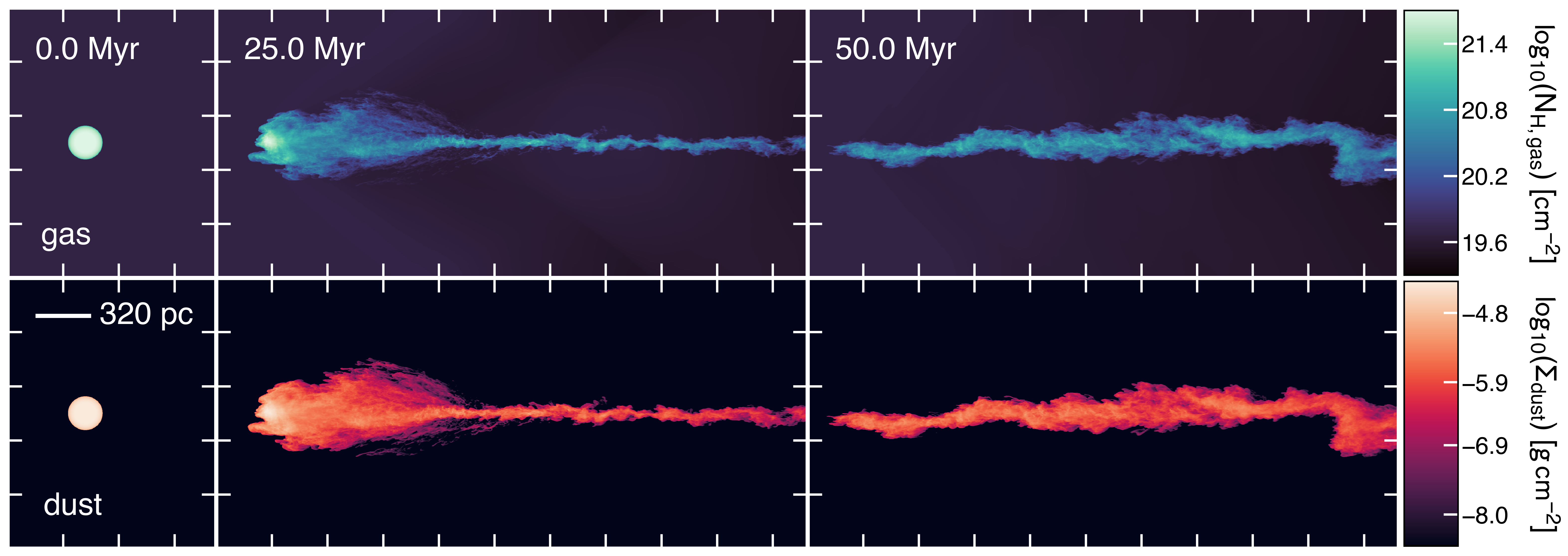}
\end{interactive}
\caption{Snapshots of the evolution of a survived cloud simulation with $a=0.1~{\mu\textrm{m}}$ dust grains. The top panel shows gas column density and the bottom panel shows dust surface density. This simulation shows the evolution of a large cloud in the slow wind model, such that $t_\mathrm{cool,minmix}\ll t_\mathrm{shear}$. Mixed phase gas forms as the wind disrupts the cloud, which quickly cools and streams directly onto a long tail behind the cloud, rather than being heated into the wind. Throughout the simulation, the distributions of dust and cloud material remain smooth and spatially coincident with one another. Since sputtering times in the cloud and mixed phase are long, essentially no sputtering is seen in the $\sim80~\textrm{Myr}$ ($5.6~t_\mathrm{sp,w}$) duration of the simulation. An animated version of this Figure can be found at \url{https://vimeo.com/showcase/11013652}.
\label{fig:survived-snapshots}}
\end{figure*}

\begin{figure}
\centering
\includegraphics[width=0.42\textwidth]{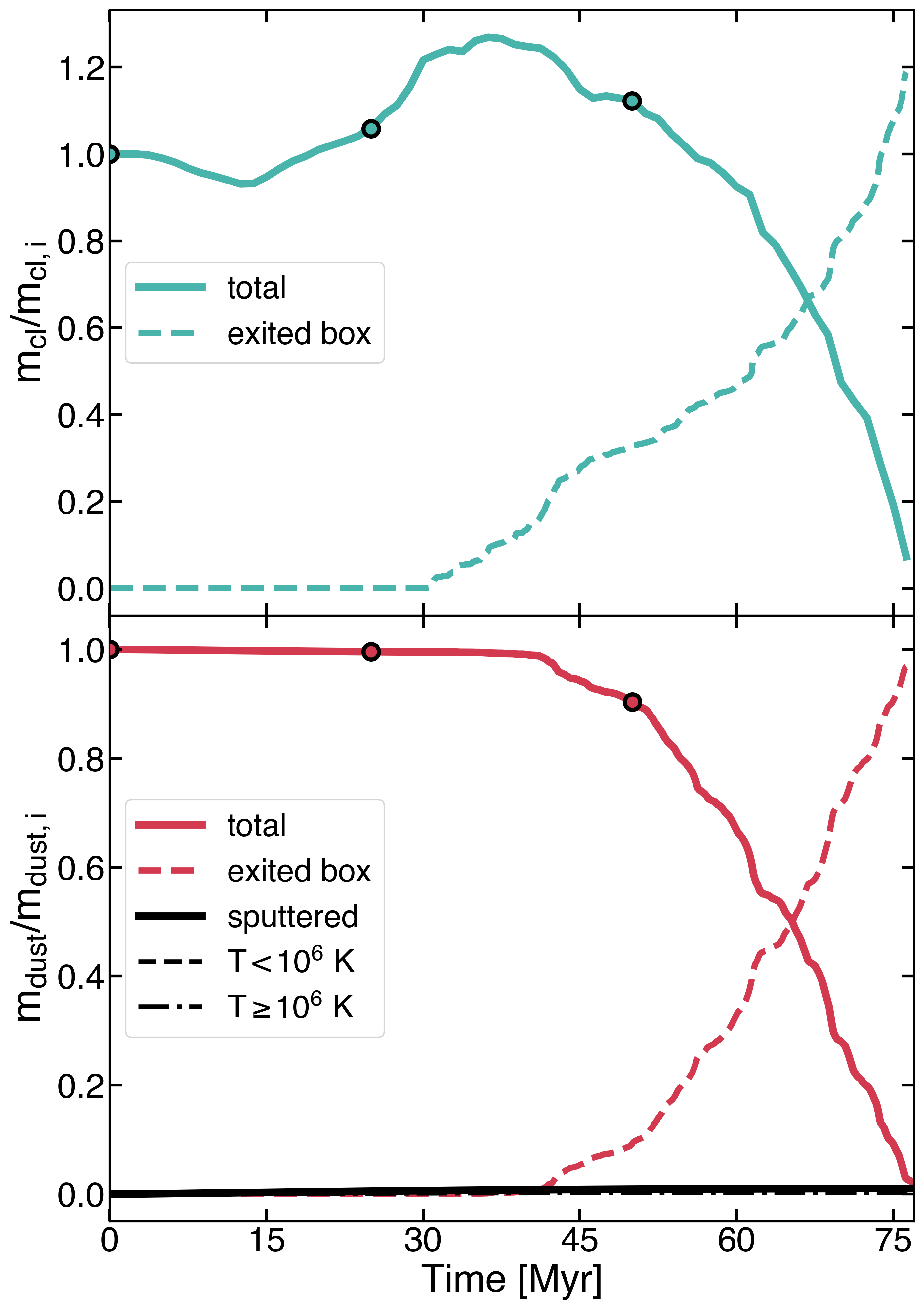}
\caption{Cloud (top) and dust (bottom) mass evolution for a survived cloud simulation with $a=0.1~{\mu\textrm{m}}$ dust grains. See Figure~\ref{fig:mass_dest} for a full description of this Figure. The initial cloud and dust masses are $6.2\times10^3~\textrm{M}_\odot$ and $63~\textrm{M}_\odot$, respectively. As the cloud evolves, it accretes rapidly cooled gas from the mixed phase and grows in mass. Essentially no sputtering is seen in the $\sim80~\textrm{Myr}$ ($5.6~t_\mathrm{sp,w}$) duration of the simulation.
\label{fig:mass_surv}}
\end{figure}

Figure~\ref{fig:destroyed-snapshots} shows three frames from this simulation: the cloud in its initial state, after it has been significantly disrupted, and after it has been largely fragmented and mixed into the wind. Figure~\ref{fig:mass_dest} shows the evolution of the simulation's cloud (top panel) and dust (bottom panel) mass. The dots in Figure~\ref{fig:mass_dest} correspond to the snapshots of gas column density (top) and dust surface density (bottom) shown in Figure~\ref{fig:destroyed-snapshots}, when the cloud mass is 100\%, 87\%, and 51\% of its initial value. The cloud mass declines monotonically throughout the $2.9~\textrm{Myr}$ run time of the simulation ($\sim19~t_\mathrm{cc}$). This is the expected behavior for a cloud of this size. The mixed phase ($n_\mathrm{mix}=0.1~\textrm{cm}^{-3}$ and $T_\mathrm{minmix}\sim9.5\times10^5~\textrm{K}$) yields $t_\mathrm{cool,minmix}=160~\textrm{kyr}$, which is long compared to the shear time ($t_\mathrm{shear}=4.9~\textrm{kyr}$), making it impossible for mixed-phase gas to cool and accrete back onto the cloud before being carried away. The rapid growth of hydrodynamic instabilities ultimately results in the cloud's destruction (see also \citealt{Schneider2017}).

The most prominent difference between the distribution of cloud and dust mass in Figure~\ref{fig:destroyed-snapshots} is a large, diffuse tail of dust that fills the volume behind the cloud. This is driven by hydrodynamic instabilities, which form as the wind sweeps over the cloud, carrying mixed-phase gas and dust away from the cloud before gas can cool back onto it \citep{Schneider2017}. The denser regions of the dust distribution trace the remaining cool gas, which has fragmented due to the rapid growth of hydrodynamic instabilities.

Once dust is removed from the cloud, it moves with the hot gas at the wind speed away from the cloud and out of the simulation volume. As shown in Figure \ref{fig:mass_dest}, after $2.9~\textrm{Myr}$ (roughly a third of the wind's sputtering time), the cloud is totally destroyed, $24\%$ of the dust mass has been sputtered. The black lines in Figure~\ref{fig:mass_dest} show that sputtering predominantly occurs in the hot phase---the dot-dashed line shows the amount of dust sputtered in $T\geq10^6~\textrm{K}$ gas, the dashed line $T<10^6~\textrm{K}$, and the solid line shows the total. The remaining dust is advected out of the simulation volume. So, relatively little sputtering is seen in this simulation, but that can largely be attributed to the short simulation duration compared to the sputtering time. The length of the box for this simulation is $0.5~\textrm{kpc}$, which is only a fraction of the sputtering distance ($r_\mathrm{sput}=5.1~\textrm{kpc}$) for this wind. In Section~\ref{subsec:hot-phase-dust}, we explore how dust will evolve once it is entrained in the hot wind. In that discussion we demonstrate that despite losing cloud shielding, roughly half of the destroyed cloud's original dust content will survive to populate the halo. This result for the fiducial case of $a=0.1~{\mu\textrm{m}}$ grains may seem surprising, given that the sputtering distance for these grains is $r_\mathrm{sput}=5.1~\textrm{kpc}$. However, the rapid decrease in the hot wind density with distance results in the conclusion that even without significant cloud shielding, long-term survival is possible for large grains in the hot phase.

\begin{figure*}
\begin{interactive}{animation}{movies/disr_movie.mp4}
\centering
\includegraphics[width=0.97\textwidth]{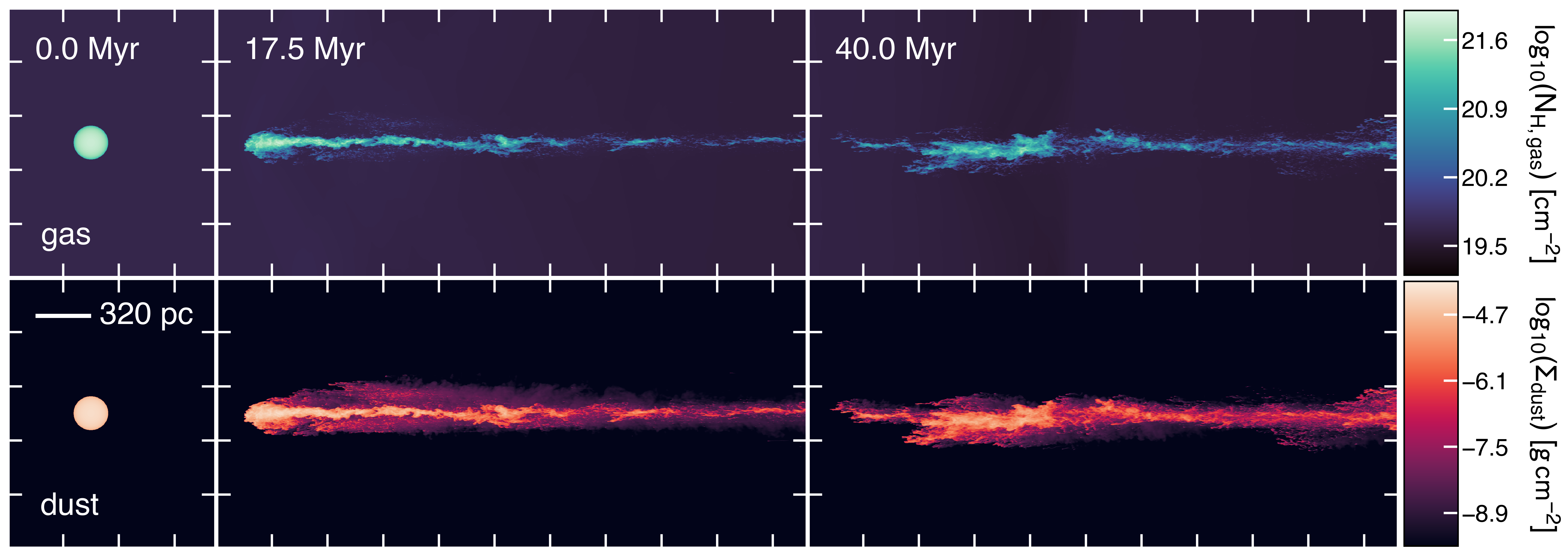}
\end{interactive}
\caption{Snapshots of the evolution of a disrupted cloud simulation with $a=0.1~{\mu\textrm{m}}$ dust grains. The top panel shows gas column density and the bottom panel shows dust surface density. This simulation uses the large cloud setup with the starburst wind model. Here, $t_\mathrm{shear}\sim t_\mathrm{cool,minmix}$, so the accretion of mixed-phase gas is less efficient than in the survived cloud simulation. Because of this, more dust gets mixed into the hot phase, resulting in the diffuse, dusty tail that forms behind the cloud's head and is not traced by cloud material. This cloud also experiences additional fragmentation, leading to an overall more disrupted morphology. After $53~\textrm{Myr}$ ($5.3~t_\mathrm{sp,w}$), $28\%$ of the initial dust mass is sputtered, primarily in the hot phase. An animated version of this Figure can be found at \url{https://vimeo.com/showcase/11013652}.
\label{fig:disrupted-snapshots}}
\end{figure*}

\subsubsection{Grain size} \label{subsubsec:dest_size}
We repeated this simulation using the same initial conditions as in Section \ref{subsec:cloud-destruction} but decreased the grain size to $a=0.01~{\mu\textrm{m}}$. Since $t_\mathrm{sp}\propto a$, the wind sputtering time is reduced to $t_\mathrm{sp,w}=1~\textrm{Myr}$, similar to the timescale for cloud destruction. Unlike in the previous simulation with larger grains, we now find that after $3.1~\textrm{Myr}$, $88\%$ of the simulation’s dust has been destroyed, with the rest carried out of the volume. $76\%$ of the sputtering occurred in $T>10^6~\textrm{K}$ gas and the remainder took place in lower-temperature gas. In this case, the surviving dust will eventually also be destroyed since the sputtering distance for 0.01 µm grains in this wind is $r_\mathrm{sp}=1~\textrm{kpc}$ and the sputtering time of the hot phase at a distance of $r=0.5~\textrm{kpc}$ is still rapid ($1.6~\textrm{Myr}$). This explains the rapid dust destruction observed in this simulation, which is perhaps more consistent with our intuition for dust destruction: the loss of cloud shielding results in nearly complete dust destruction. As a result, we conclude that in this cloud evolution scenario (with $t_\mathrm{cool,minmix}>t_\mathrm{shear}$) dust grains with radius $\lesssim0.01~{\mu\textrm{m}}$ will be completely destroyed.

\subsection{Survived clouds} \label{subsec:cloud-survival}
Figure~\ref{fig:survived-snapshots} shows snapshots of the evolution of gas and dust in our survived cloud simulation, the parameters of which are given in the bottom section of Table~\ref{tab:sims}. This simulation models a large cloud evolving in the slow wind model.  Because $t_\mathrm{sp}$ is roughly constant for temperatures above $T_\mathrm{sp}=2\times10^6~\textrm{K}$, the sputtering time of this wind is close to that of the starburst wind model---$t_\mathrm{sp,w}=14~\textrm{Myr}$---even though this wind is an order of magnitude cooler. The mixed-phase parameters for this cloud are $n_\mathrm{mix}=0.3~\textrm{cm}^{-3}$ and $T_\mathrm{minmix}=3\times10^5~\textrm{K}$, resulting in $t_\mathrm{cool,minmix}=41~\textrm{kyr}$. This is short compared to the shear time, which is $\sim200~\textrm{kyr}$. As a result, the cloud efficiently absorbs gas from the mixed phase, enabling long-term survival and growth (as shown in the top panel of Figure~\ref{fig:mass_surv}). After several $t_\mathrm{cc}$, cloud material is extended throughout a long tail that forms behind its head, spanning $\sim10~\textrm{kpc}$. Throughout the simulation, the distribution of dust and cloud material is quite smooth and continuous, as shown in Figure~\ref{fig:survived-snapshots}---very little dust enters the hot wind. This demonstrates that all of the mixed-phase gas ends up condensing onto the cloud's tail instead of mixing into the hot phase, which results in very efficient shielding of dust from the hot phase. Figure~\ref{fig:mass_surv} shows that essentially no dust is sputtered throughout the $\sim80~\textrm{Myr}$ ($5.6~t_\mathrm{sp,w}$) duration of the simulation.

\subsubsection{Grain size}

We repeated this simulation for $0.01~{\mu\textrm{m}}$ radius dust grains (shown in Table~\ref{tab:sims}) and found similar results. Since no gas is transferred to the hot phase, dust spends most of its time shielded by the cloud where the sputtering time is long ($12-0.37~\textrm{Gyr}$, between the cloud's initial conditions and its final, atomic-phase state). The mixed-phase sputtering time for $a=0.01~{\mu\textrm{m}}$ grains in this simulation is also quite long ($t_\mathrm{sp,minmix}\sim70~\textrm{Myr}$), so the brief mixed-phase exposure that dust experiences has little effect. Overall, $8\%$ of the initial dust mass is sputtered in this simulation (after $58~\textrm{Myr}$, or $41~t_\mathrm{sp,w}$), almost entirely in $T<10^6~\textrm{K}$ gas. We ran an additional simulation of this cloud with $a=10~\textup{\AA}$ radius grains. We found that sputtering within the cloud and mixed-phase becomes much more significant for grains of this size, with $53\%$ of the $10~\textup{\AA}$ grains destroyed all within $T<10^6~\textrm{K}$ gas. We discuss the implication of these results in Section~\ref{subsec:grainsize}.

\begin{figure}
\centering
\includegraphics[width=0.42\textwidth]{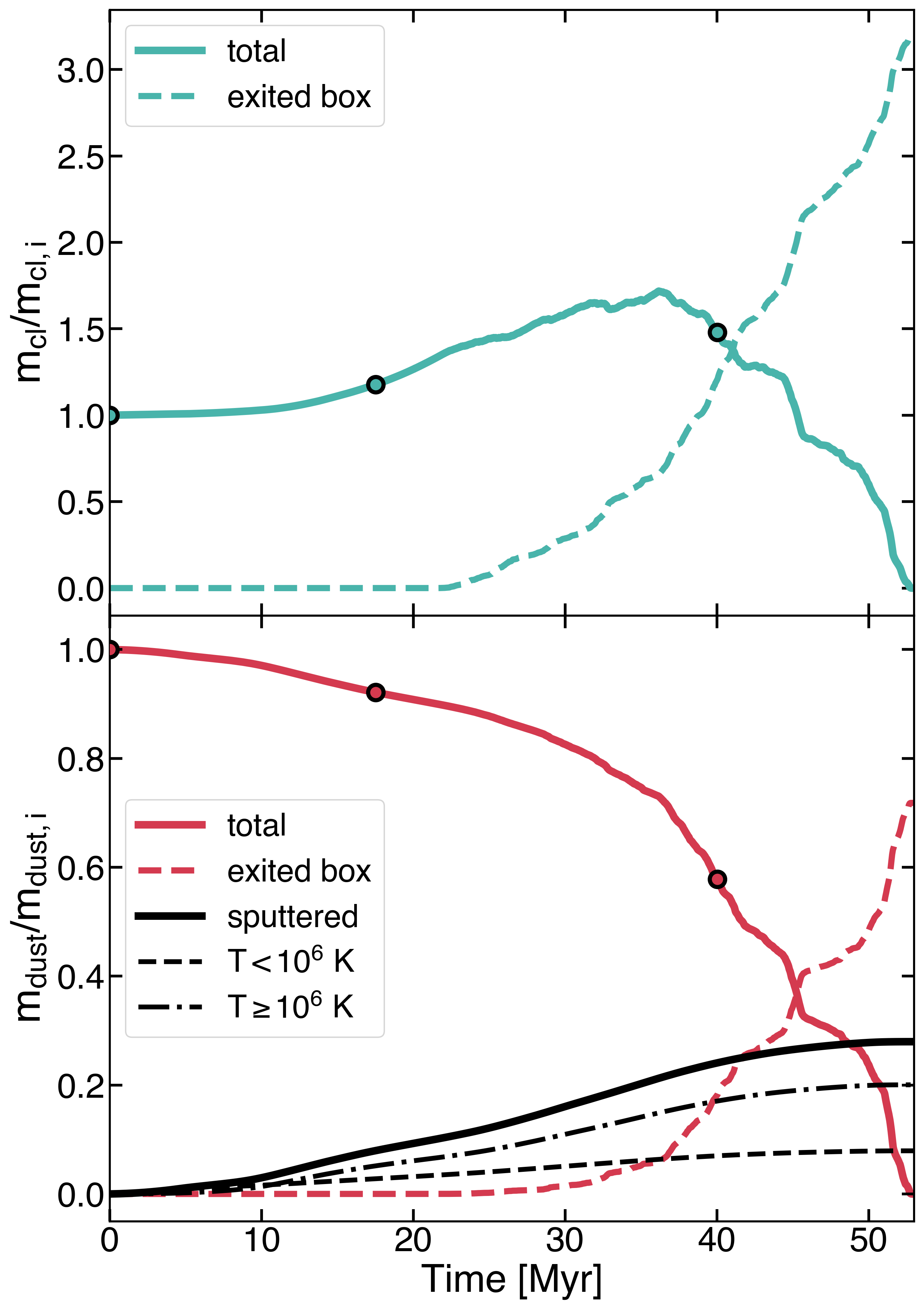}
\caption{Cloud (top) and dust (bottom) mass evolution for a disrupted cloud simulation with $a=0.1~{\mu\textrm{m}}$ dust grains. See Figure~\ref{fig:mass_dest} for a full description of this Figure. The initial cloud and dust masses are $6.2\times10^3~\textrm{M}_\odot$ and $63~\textrm{M}_\odot$, respectively. This cloud exhibits signs of growth, with its total mass increasing slightly before significant cloud material begins to exit the simulation volume. After $53~\textrm{Myr}$ ($5.3~t_\mathrm{sp,w}$), $28\%$ of the initial dust mass is destroyed, mostly in the hot phase.
\label{fig:mass_disr}}
\end{figure}

\begin{figure}
\centering
\includegraphics[width=0.42\textwidth]{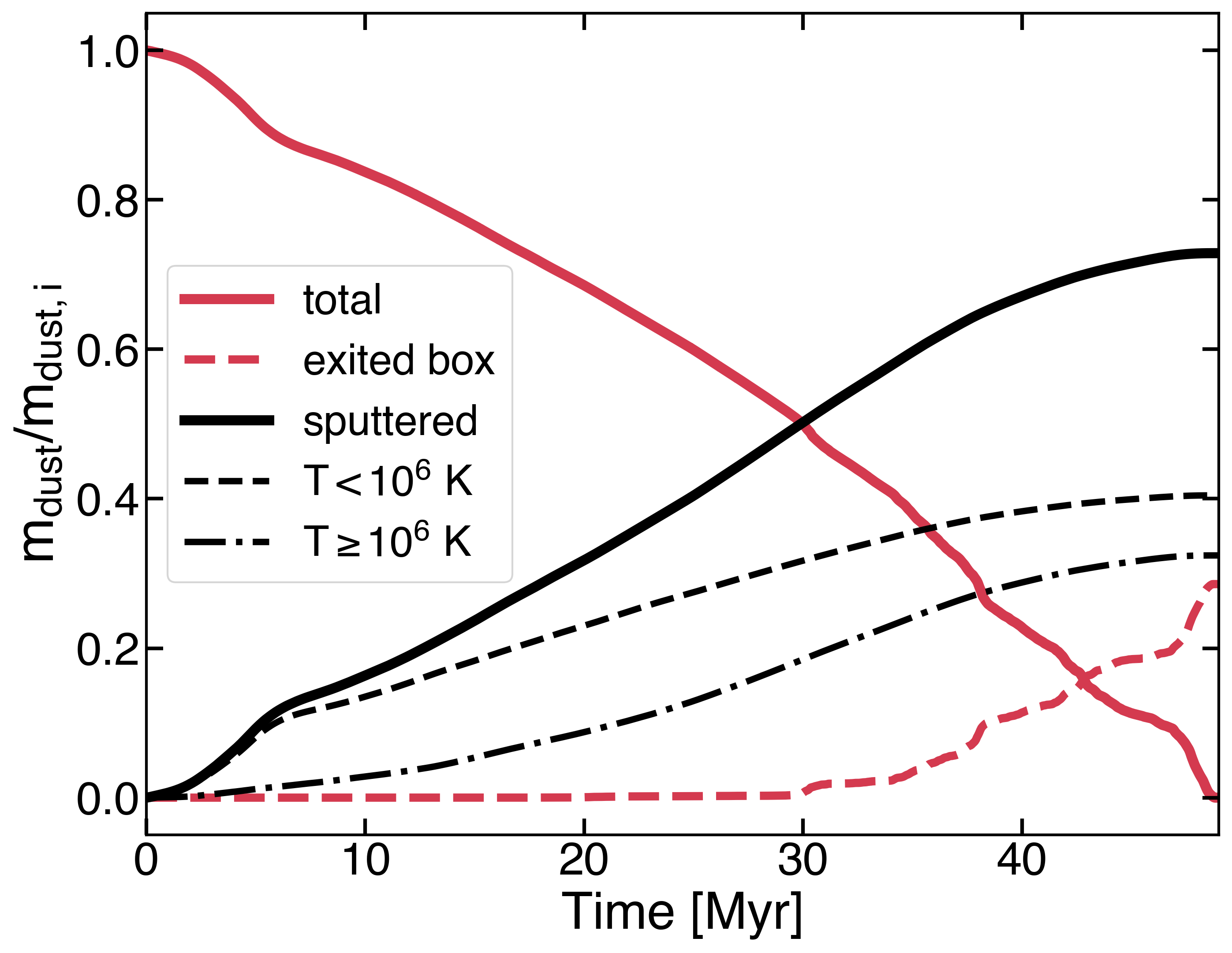}
\caption{Dust mass evolution for the same simulation setup as described in Figure~\ref{fig:mass_disr}, but with $a=0.01~{\mu\textrm{m}}$ dust grains. After $49~\textrm{Myr}$ ($49~t_\mathrm{sp,w}$), $73\%$ of the initial dust mass is destroyed, with a higher fraction of destruction occurring in $T<10^6~\textrm{K}$ gas. This is due to the relatively short mixed-phase sputtering time at this grain size---$t_\mathrm{sp,minmix}\sim250~\textrm{kyr}$. This is comparable to $t_\mathrm{cool,minmix}$, so mixed-phase sputtering becomes efficient.
\label{fig:mass_disr_small}}
\end{figure}

\subsection{Disrupted clouds} \label{subsec:cloud-disruption}

To understand how $a=0.1~{\mu\textrm{m}}$ dust evolves in a long-lived, but more fragmented cloud, we simulated a large cloud in our starburst wind model (parameters shown in the middle section of Table~\ref{tab:sims}). In this simulation, $t_\mathrm{cool,minmix}\sim t_\mathrm{shear}$. Figure \ref{fig:disrupted-snapshots} shows snapshots of the cloud evolution, and Figure \ref{fig:mass_disr} shows the cloud and dust evolution. The mixed phase gas in this simulation has a density of $n_\mathrm{mix}=0.3~\mathrm{cm}^{-3}$ and temperature $T_\mathrm{minmix}=9.5\times10^5~\textrm{K}$, resulting in $t_\mathrm{cool,minmix}\sim160~\textrm{kyr}$. The shear time is slightly shorter than this: $t_\mathrm{shear}\sim98~\textrm{kyr}$, which should result in cloud survival according to Eq.~\ref{eq:crit}, but the absorption of mixed-phase material is less efficient than in our survived cloud simulations. This is illustrated in Figure~\ref{fig:disrupted-snapshots}, where we see the cloud has formed a long tail from the accretion of mixed-phase gas, similar to the survived cloud (see Figure~\ref{fig:survived-snapshots}). However, the disrupted cloud has a more fragmented distribution of gas and dust in its tail, particularly at early times. In the later snapshots, a diffuse haze of dust can be seen surrounding the tail in areas not traced by cloud material. This haze originates from dust in the mixed phase that was unable to condense back onto the cloud.

The inefficient re-accretion of mixed-phase gas in the disrupted cloud leads to enhanced dust sputtering because less dust stays inside of the cloud. This can be seen in the bottom panel of Figure \ref{fig:mass_disr}, which shows that $28\%$ of the initial dust mass is destroyed and that this predominantly takes place in the hot phase. Sputtering in the mixed phase itself is less efficient since the mixed-phase sputtering time is significantly longer than the cooling time ($t_\mathrm{sp,minmix}\sim16~t_\mathrm{cool,minmix}$).

While the disrupted cloud's less efficient accretion of mixed-phase gas helps explain the relatively fragmented nature of the tail, the initial thermal instability of the disrupted cloud may also contribute. In more detail, the disrupted cloud is initialized at $3\times10^4~\textrm{K}$ and immediately cools to $10^4~\textrm{K}$. \citet{Gronke2020b} suggest that the thermal instability may cause the clouds to break up into a mist of cloudlets. While this may increase the amount of mixing (and consequently the dust mass evolution) compared to a cloud initialized in temperature equilibrium, this does not affect our conclusions. After all, we are considering idealized uniform clouds, which have different mixing properties from more realistic clouds with turbulent density distributions \citep[e.g.][]{Schneider2017,Banda-Barragan2019,Gronke2020a}.

\subsubsection{Grain size}
We repeated this simulation for $0.01~{\mu\textrm{m}}$ grains, shown in Table~\ref{tab:sims}. The cloud evolution is the same as described in Section~\ref{subsec:cloud-disruption}. The evolution of this simulation's dust mass is shown in Figure \ref{fig:mass_disr_small}. Here, significant dust destruction occurs---$73\%$ of the initial dust mass is destroyed, with $40\%$ occurring in $T<10^6~\textrm{K}$ gas and $33\%$ in the hot phase. In this scenario, the mixed-phase sputtering time and cooling time are close to one another---$t_\mathrm{sp,minmix}\sim1.7~t_\mathrm{cool,minmix}$. As a result, significant sputtering can occur in the mixed phase before gas can cool and condense onto the cloud. Overall, grains of this size and smaller are unlikely to survive in disrupted clouds.

\subsection{Summary}
In this Section, we described the evolution of dust grains of varying sizes in the three regimes of cloud survival. We find that cloud evolution (determined by $t_\mathrm{cool,minmix}/t_\mathrm{shear}$) has the most significant effect on dust survival, followed by dust grain size. Clouds in the survival regime are very effective at shielding dust from the hot and mixed phases since nearly all of the disrupted cloud material efficiently condenses back onto the cloud. $0.1~{\mu\textrm{m}}$ and $0.01~{\mu\textrm{m}}$ radius grains experienced little to no sputtering in survived clouds and $10~\textup{\AA}$ grains exhibited partial survival. Disrupted clouds result in slightly enhanced amounts of sputtering for $0.1~{\mu\textrm{m}}$ radius grains since some mixed-phase gas is unable to condense back onto the cloud, resulting in increased hot-phase exposure for dust. However, a majority of dust still survives, with only $28\%$ of the initial dust mass sputtered in the simulation. For $0.01~{\mu\textrm{m}}$ radius grains, the combination of hot-phase exposure and shorter mixed-phase sputtering times was significantly detrimental, and $70\%$ of the initial dust mass was destroyed. Destroyed clouds result in the most dust destruction, with $50\%$ destruction for $0.1~{\mu\textrm{m}}$ radius grains (as we will demonstrate in Section \ref{subsec:hot-phase-dust}) and total destruction for $0.01~{\mu\textrm{m}}$ radius grains and smaller. In short, survived, disrupted, and destroyed clouds are all capable of transporting $a=0.1~{\mu\textrm{m}}$ dust to the halo, with varying percentages of dust survival, but only clouds that survive are capable of transporting grains smaller than this to these regions.

\section{Discussion} \label{sec:discussion}

\subsection{The fate of hot-phase dust} \label{subsec:hot-phase-dust}

The simulation discussed in Section \ref{subsec:cloud-destruction} represents the least amenable scenario to dust survival: a small cloud that is quickly destroyed by a hot, rapid wind. In this case, after roughly $3~\textrm{Myr}$ ($t/t_\mathrm{sp,w}=0.29$), the cloud is destroyed and the remaining dust is carried out of the simulation volume. Thus, some dust is destroyed in the simulation ($24\%$ of the initial dust mass), but most of the dust ``survives" because we lose our ability to track it. In this Section, we use semi-analytic calculations to address the fate of this hot-phase dust once it moves beyond the explicitly tracked simulation domain.

In Section~\ref{subsubsec:hotphase}, we used an analytic argument to show that it may be possible for dust to survive even when fully exposed to the wind, due to the rapid decline in hot phase density and temperature (as seen in the CGOLS datapoints in Figure \ref{fig:t_sp}). To fully investigate this scenario, we solved the sputtering equation (Eq.~\ref{eq:sput-model}) for a self-consistent outflow profile to determine how dust would evolve after it entered the hot phase and was fully exposed to the wind. To do this, we assumed that the dust was moving at the hot wind speed and that the density and temperature declined with distance, following the CGOLS hot-phase profiles. At each time step, we updated the dust position assuming it was moving at $v_\mathrm{w}=10^3~{\textrm{km}\,\textrm{s}^{-1}}$, and used the CGOLS hot phase profiles at that position to update the sputtering time.

\begin{figure}
\centering
\includegraphics[width=0.42\textwidth]{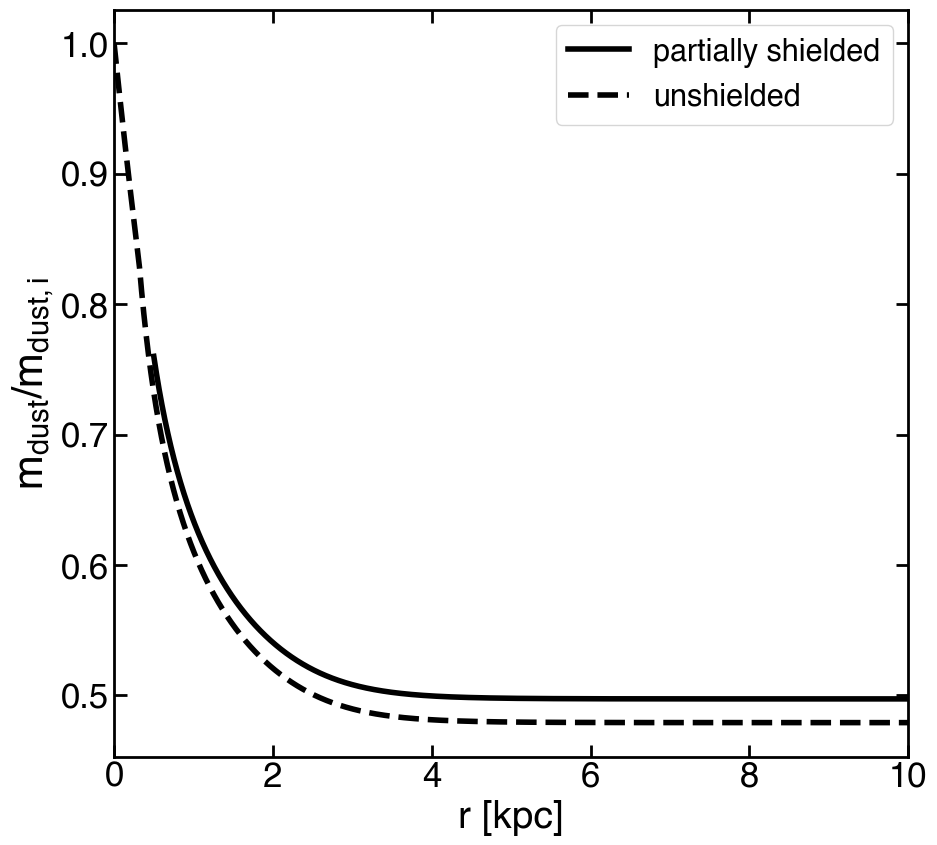}
\caption{Evolution of the dust mass with distance due to sputtering for dust moving at $10^3~{\textrm{km}\,\textrm{s}^{-1}}$ in a hot wind. The partially shielded curve represents a continuation of the simulation discussed in Section \ref{subsec:cloud-destruction}, beginning at a distance of $r=0.5~\textrm{kpc}$ from the base of the outflow. The unshielded curve shows what happens to dust that is never inside of a cloud, and travels from the base of the wind fully exposed to the hot phase. As dust moves outward, the sputtering timescale evolves according to analytic fits to the CGOLS hot phase profiles. A total of $50\%$ of the initial dust density is sputtered in the partially shielded case. $48\%$ of the initial dust density is sputtered in the unshielded case.
\label{fig:single_cell}}
\end{figure}

We solved this model out to a distance of $10~\textrm{kpc}$ using analytic fits to the CGOLS hot phase profiles, $n\propto r^f$ and $T\propto r^{f(\gamma-1)}$ \citep{Schneider2020}, where $f=0.05\,r-1.08$ and $\gamma$ is the adiabatic index of the gas, taken to be 5/2. We explored two versions of this model. The ``partially shielded" case corresponds to a continuation of the simulation in Section \ref{subsec:cloud-destruction}, where $76\%$ of the ($a=0.1~{\mu\textrm{m}}$) dust is transported a distance of roughly $r=0.5~\textrm{kpc}$ shielded by the cloud before it is fully transferred into the wind. The ``unshielded" case begins at $r=0~\textrm{kpc}$, representing the maximally destructive scenario with zero cloud shielding. In the unshielded case, dust is assumed to traverse the hottest, densest part of the outflow completely exposed to the hot phase.

\begin{figure*}
\centering
\includegraphics[width=\textwidth]{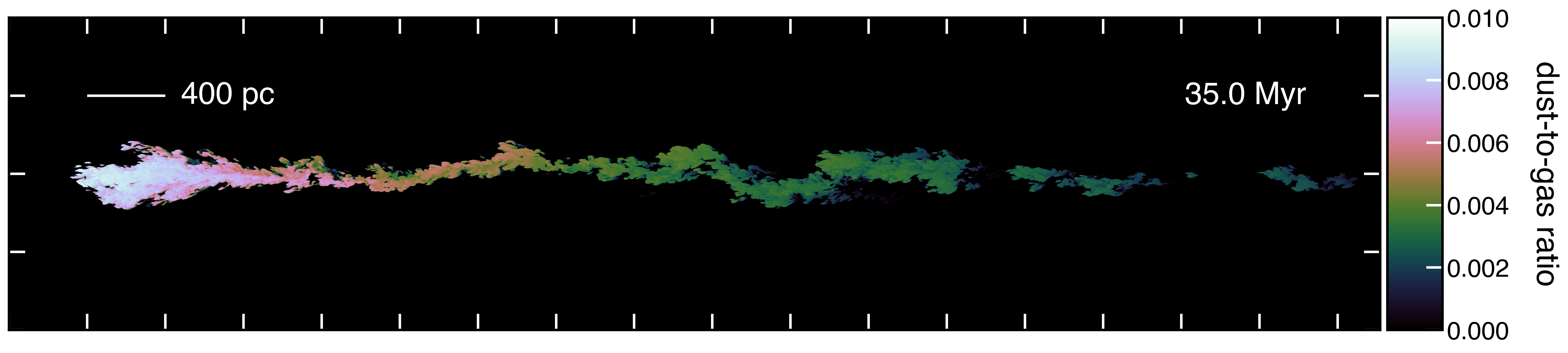}
\caption{Slice of a survived cloud simulation showing the variation in dust-to-gas ratio throughout the cloud (over a distance of $7.0~\textrm{kpc}$). The head of the cloud is at roughly the initial dust-to-gas ratio (0.01), but the cloud becomes less dusty with distance from the head, declining by roughly an order of magnitude. This results from the condensation of the mixed phase, which is composed of dust-rich gas from the cloud and dust-free gas from the hot phase.
\label{fig:dtg_slice}}
\end{figure*}

\begin{figure}
\centering
\includegraphics[width=0.42\textwidth]{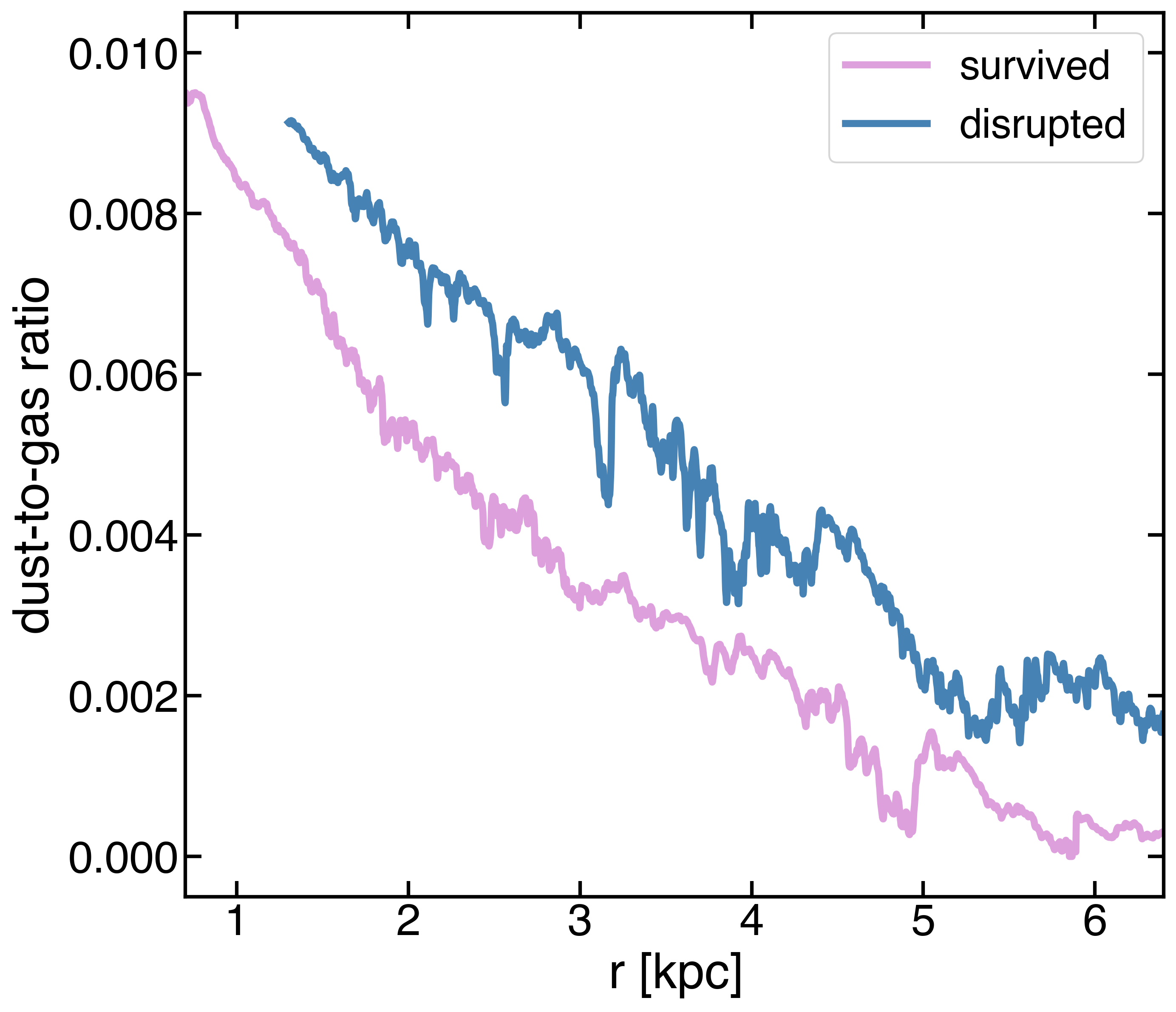}
\caption{Dust-to-gas ratio of cool-phase gas as a function of distance for a snapshot at $\sim20~\textrm{Myr}$ of a disrupted (blue line) and survived (pink line) cloud. The initial dust-to-gas ratio of each simulation is 0.01. The decrease in dust-to-gas ratio with distance is a reflection of the constant dust mass as the clouds accrete gas from the wind.  
\label{fig:dtg}}
\end{figure}

Figure \ref{fig:single_cell} shows the evolution of the dust mass as it is carried outward. There is a sharp decline in dust mass while dust is still relatively close to the base of the outflow ($r\lesssim 2~\textrm{kpc}$), but beyond this point, the remaining amount of dust is nearly constant. In the partially shielded case, $m_\mathrm{dust}/m_\mathrm{dust,i}=0.50$. This means that half of the cloud's initial dust mass ($0.077~\textrm{M}_\odot$) is sputtered by the time it reaches the halo. To understand the effect that partial cloud shielding has on the total amount of dust survival, we turn to the case of unshielded dust. Below $r=0.5~\textrm{kpc}$, the analytic functions are a poor fit to the CGOLS data (see Figure 6 of \citealt{Schneider2020}), so we assume they are constant at values of $n(r=0.5~\textrm{kpc})$ and $T(r=0.5~\textrm{kpc})$ for $r<0.5~\textrm{kpc}$. The result of this model is also shown in Figure \ref{fig:single_cell}. In this case, $48\%$ of the dust survives. These results imply that cloud shielding has essentially no effect on dust survival. Note that these results are for $a=0.1~{\mu\textrm{m}}$ grains. As discussed in Section \ref{subsubsec:dest_size}, $a=0.01~{\mu\textrm{m}}$ grains are completely destroyed in the case of cloud destruction.

\subsection{Dust-to-gas ratio} \label{subsec:dtg}
We have shown in Sections~\ref{subsec:cloud-survival} and \ref{subsec:cloud-disruption} that dusty clouds can be long-lived if rapid cooling of the mixed phase transfers mass and momentum to the cool phase. Clouds that meet this condition form long tails of condensed mixed gas. Because these tails form from a mixture of dust-rich cloud gas and dust-poor hot-phase gas, we expect that they will have a lower dust-to-gas ratio than the initial cloud. Figure~\ref{fig:dtg_slice} illustrates this. Here, we see a slice of the survived cloud described in Section~\ref{subsec:cloud-survival} roughly halfway through the simulation. The physical extent of this slice is $7.0~\textrm{kpc}$ long and $1.6~\textrm{kpc}$ high. At this point, the initial cloud is still being disrupted, but a long tail has formed behind it. Thus, the head of the cloud still has roughly the same dust-to-gas ratio as its initial value (0.01). However, the tail that extends behind the head of the cloud is much less dusty---the dust-to-gas ratio furthest from the head is almost an order of magnitude lower. This result holds for both our survived and disrupted cloud simulations, as shown in Figure~\ref{fig:dtg} since both clouds must condense gas out of the hot wind to survive. This shows both clouds at the same point in time ($\sim20~\textrm{Myr}$ into the simulation). It is worth noting that in these simulations very little dust is sputtered, and thus the changing dust-to-gas ratio is solely a reflection of the clouds' accretion of dust-free hot gas. In fact, the disrupted cloud simulation experiences \emph{more} sputtering, but has, on average, a higher dust-to-gas ratio, because gas accretion is less efficient. Over time, as the cloud continues to mix with the hot phase, the dust-to-gas ratio becomes more homogeneous throughout the cloud.

\subsection{Halo dust masses} \label{subsec:halo_dust}

In this Section, we use the results of our simulations combined with an analysis of the population of clouds in an outflow to estimate the total amount of $a=0.1~{\mu\textrm{m}}$ dust that a typical wind in a starburst galaxy could transfer to the halo. Using CGOLS data, we have catalogued the cloud properties in the outflow of a $\textrm{M}_\mathrm{star}=10^{10}~\textrm{M}_\odot$ galaxy with a star formation rate of $20~\textrm{M}_\odot\,\textrm{yr}^{-1}$ (Warren et al., \emph{in prep}). This catalogue contains the total number of clouds, along with their masses and sizes. The cloud masses in this distribution follow $\textrm{d}N/\textrm{d}M\propto M^{-2}$, similar to results found in other theoretical studies of cloud distributions (e.g. \citealt{Tan2024}, Figure 16). We divide the cloud distribution into three bins based on cloud radius, of destroyed, disrupted, and survived clouds, and apply the dust survival percentages found in this work (for $0.1~{\mu\textrm{m}}$ radius grains) to estimate the total survived dust mass in each bin. Cloud radii in the catlogue range from $\sim5-300~\textrm{pc}$, with a combined mass of $\sim10^7~\textrm{M}_\odot$. 

To divide this catalogue into three regimes of dust evolution, we estimate the critical radius, $r_\mathrm{crit}$, for cloud survival in the CGOLS wind. We use a pressure of $\log_{10}(p/k_\mathrm{B})=5$, a temperature of $T_\mathrm{w}=3\times10^7~\textrm{K}$, a velocity of $v_\mathrm{w}=10^3~{\textrm{km}\,\textrm{s}^{-1}}$, and a density contrast of $\chi=10^3$, which are roughly the values of the wind profiles near the base of the outflow. This results in $r_\mathrm{crit}\sim70~\textrm{pc}$ (see Eq. 3 of \citet{Abruzzo2023} for details on calculating $r_\mathrm{crit}$).

Clouds below this radius are in the destruction regime and account for roughly $65\%$ of the cloud mass in the distribution. As shown in Section~\ref{subsec:hot-phase-dust}, approximately $50\%$ of the dust will be destroyed in this case. We take $t_\mathrm{cool,minmix}/t_\mathrm{shear}=2$ to roughly mark the cutoff between disrupted and survived clouds, which corresponds to a cloud radius of $245~\textrm{pc}$. Clouds below this size, but above $r_\mathrm{crit}$, are in the disrupted regime and account for roughly $25\%$ of the cloud mass in the distribution. Our simulations show that clouds in this regime exhibit $28\%$ dust destruction. Finally, clouds above this radius are in the survival regime and account for $10\%$ of the total cloud mass. We conclude that these clouds safely transport all of their initial dust mass. 

Combining these totals, we estimate that approximately $60\%$ of the initial dust mass will survive its trip to the halo in this kind of outflow. This estimate represents the total fraction of $a=0.1~{\mu\textrm{m}}$ dust that would get carried to the halo by a typical starburst galaxy in this mass regime. For the specific CGOLS galaxy considered here, the cool gas mass loading factor (outflow rate relative to star formation rate) at $1~\textrm{kpc}$ above the disk is roughly $\eta_\mathrm{g}=0.1$ (\citealt{Schneider2024}, Figure 7), implying a cool gas outflow rate of $\eta_\mathrm{g}\times \textrm{SFR}=2~{\textrm{M}_\odot\,\textrm{yr}^{-1}}$. Given an ISM dust-to-gas ratio of 0.01 and a dust survival fraction of 0.6, this would result in a ``dust outflow rate" of $\sim 0.01~\textrm{M}_\odot\,\textrm{yr}^{-1}$. In future works, we will apply this dust model to full-galaxy simulations, where we will be able to test this prediction given a full cloud distribution, enabling more accurate estimates of predicted halo dust mass.

\subsection{Grain size distribution} \label{subsec:grainsize}
We carried out simulations of $0.1~{\mu\textrm{m}}$ and $0.01~{\mu\textrm{m}}$ radius grains in each case of cloud survival. We find that grains of radius $0.01~{{\mu}\textrm{m}}$ and smaller are unable to withstand hot phase exposure, which occurs in the case of cloud destruction and disruption. In clouds that are destroyed, small grains are fully exposed to the hot phase and are quickly sputtered. In the case of disrupted clouds, some dust ends up in the hot phase due to the inefficiency of cooling, resulting in some small grain destruction (as is shown in Figure~\ref{fig:mass_disr}). Mixed-phase exposure is also detrimental to $0.01~{\mu\textrm{m}}$ grains in disrupted clouds since $t_\mathrm{sp,minmix}$ is comparable to the time dust spends in this phase. This means that the only scenario in which $0.01~{\mu\textrm{m}}$ grains aren't subject to high amounts of sputtering is in clouds that survive, where time spent in the mixed phase is brief (due to rapid cooling and condensation) and hot phase exposure does not occur.

\begin{figure}
\centering
\includegraphics[width=0.42\textwidth]{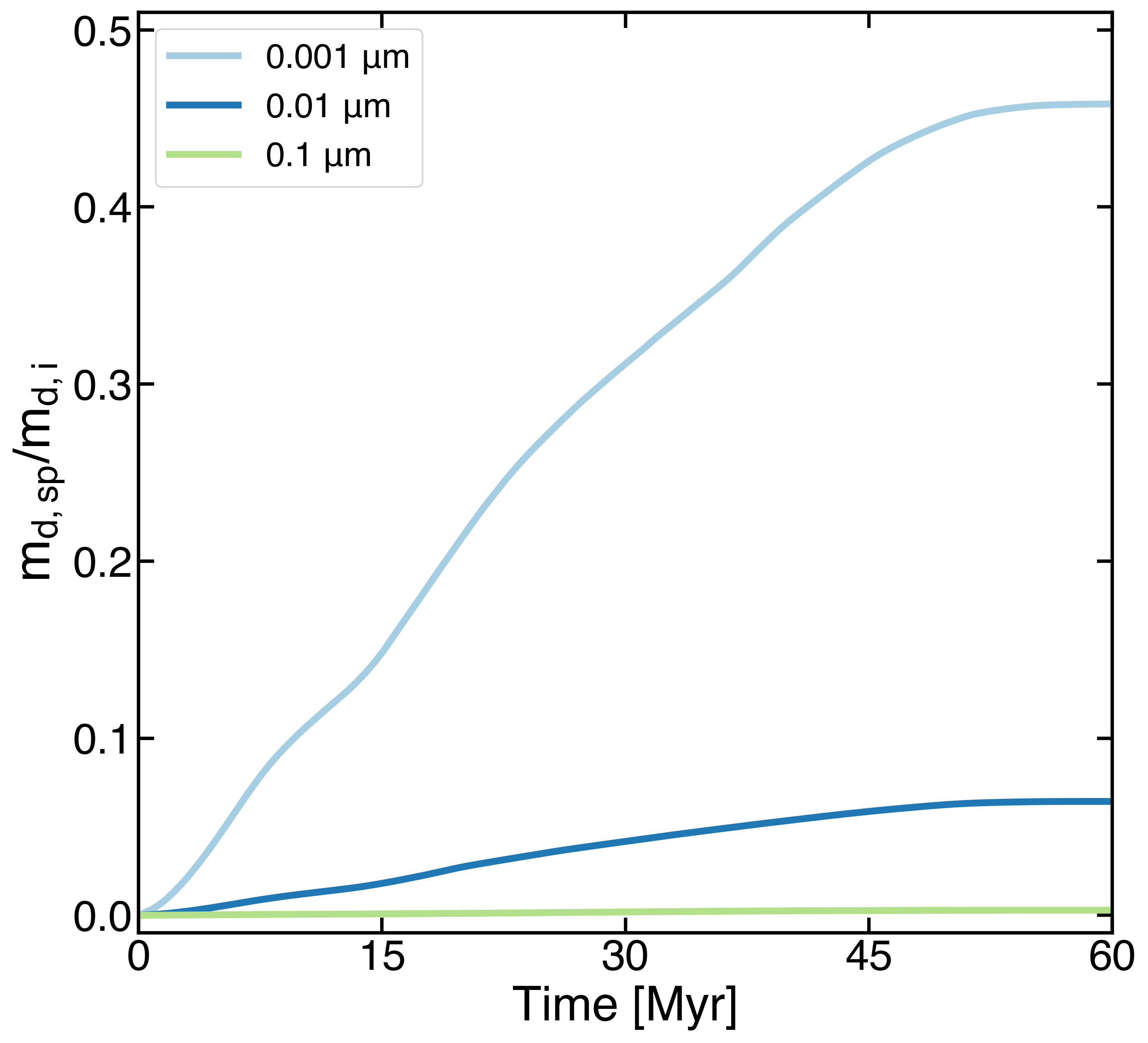}
\caption{The fraction of dust destruction seen for varying grain sizes in survived cloud simulations (where $t_\mathrm{cool,minmix}\ll t_\mathrm{shear}$, described in Section~\ref{subsec:cloud-survival}). Essentially all sputtering occurs in $T<10^6~\textrm{K}$ gas. \label{fig:sputter_frac}}
\end{figure}

We ran additional simulations of our survived cloud with $10~\textup{\AA}$ radius grains to determine the smallest grain size that could be long-lived in an outflow. We found that slightly more than half of the $10~\textup{\AA}$ grains survived. Cool and mixed-phase sputtering both contribute to the decrease in dust survival since the sputtering time of atomic-phase cloud material ($T\sim10^4~\textrm{K}$ and $n\sim1~\textrm{cm}^{-3}$) is only $37~\textrm{Myr}$. It should be noted that Eq.~\ref{eq:sput-timescale} may overestimate dust lifetimes for very small dust grains \citep{Micelotta2010}, so this survival fraction is likely an upper limit. The fraction of dust destruction for all three grain sizes in the survived cloud is shown in Figure~\ref{fig:sputter_frac}.

In summary, outflows with cool ($\sim10^4~\textrm{K}$) clouds that have $t_\mathrm{cool,minmix}\ll t_\mathrm{shear}$ can transport grains down to $\sim10~\textup{\AA}$ in radius. Clouds with $t_\mathrm{cool,minmix}\gtrsim t_\mathrm{shear}$ may exhibit a shift in the grain size distribution toward larger grains as a result of sputtering, since dust smaller than $0.1~{\mu\textrm{m}}$ in radius cannot survive. It is difficult to predict what the overall grain size distribution in these outflows would be, however, because the shattering of large grains within clouds may repopulate smaller grain sizes. We discuss the potential importance of shattering in outflows in Section \ref{subsec:mechanisms}.

\subsection{Other dust evolution mechanisms} \label{subsec:mechanisms}

Our dust model treats dust as a fluid fully dynamically coupled to gas and subject to destruction through sputtering. In many circumstances, this is a good approximation, as gas ions and charged dust grains are frequently spatially coincident due to their mutual attraction to magnetic field lines. Gas moves along these field lines and dust (which gyrates around them at a distance of the Larmor radius) accelerates with its bulk motion by drag forces, making the dynamical coupling of gas and dust a good approximation \cite{Draine2011}. Sputtering, in this case, is driven by gas thermal motion (as opposed to gas-grain relative velocities). However, thermal sputtering is not the only mechanism that affects dust. Dust grains can shatter in grain-grain collisions, grow through grain-grain coagulation, and accrete metals from the gas phase. Grains can also become dynamically decoupled from gas due to turbulence and shocks. Ignoring accretion and coagulation should have little effect on our results since these mechanisms are only efficient in very cold and dense regions of the ISM (see, for example, \citealt{Sembach1996, Priestley2021, Chokshi1993}). Shattering and nonthermal sputtering, however, may affect dust survival, particularly if gas-grain decoupling is significant. We discuss the potential for these mechanisms to impact our results in the following Sections.

\subsubsection{Nonthermal sputtering}

In dust evolution modeling, a somewhat artificial distinction is made between thermal and nonthermal (or inertial) sputtering. Thermal sputtering results from gas-grain collisions driven by gas thermal motions. Here, sputtering depends on temperature, which increases with gas velocity dispersion, resulting in more frequent collisions. Nonthermal sputtering refers to sputtering caused by collisions due to nonzero gas-grain relative bulk velocities. Our simulations do not account for nonthermal sputtering since we assume gas and dust are fully dynamically coupled. Although shocks in winds may be able to briefly decouple the motions of gas and dust, we have analyzed the nonthermal sputtering and gas drag times in these wind conditions and determined that dust will usually recouple on timescales shorter than the nonthermal sputtering timescale (see Appendix \ref{app:nonth}). Given this, we expect that ignoring non-thermal sputtering and decoupled dust dynamics should have little effect on the results of this study.

\subsubsection{Shattering}

Shattering occurs when there are non-zero grain-grain relative velocities, resulting in collisions between dust grains that lead to their fragmentation. Generally, this process converts large ($a\gtrsim0.1~{\mu\textrm{m}}$) grains to small ($a\lesssim0.01~{\mu\textrm{m}}$) grains and is believed to be a significant source of small grains, such as polycyclic aromatic hydrocarbons, in the Universe. Shattering can be induced by gas turbulence \citep{Hirashita2009} as well as shocks \citep{Jones1996}. Our simulations do not include shattering, since we assume a single grain size and do not model dust dynamics independent of the gas. However, it is possible that shattering could affect the total dust survival fraction in winds. Analytic studies have shown that turbulence-induced shattering in cool gas ($n_\mathrm{H}=0.1~\textrm{cm}^{-3}$, $T_\mathrm{gas}=10^4~\textrm{K}$) can occur on timescales of $\sim10^8~\textrm{yr}$ \citep{Hirashita2021}, which could reduce the overall amount of grain survival in survived and disrupted clouds. Additionally, the extent to which shock-induced shattering may affect grains in outflows is currently unclear. If efficient, shock-induced shattering could serve to reduce the overall dust survival fraction in disrupted and destroyed clouds, since very small grains $a<0.01~{\mu\textrm{m}}$ are subject to destruction in these scenarios.

Observations are beginning to shed light on these questions. For example, recent observations of dust emission in the outflow of nearby edge-on Milky Way analog galaxy NGC 891 have shown evidence of an increasing fraction of small grains with distance from the disk \citep{Katsioli2023}, which may be an indication that shattering is taking place in the outflow. More work is needed to fully understand the source and evolution of small grains in outflows. We plan to address these uncertainties in future work by including models for the grain size distribution and dust dynamics in our simulations.

\subsection{Comparison with other work}

While dust destruction in outflows has not been thoroughly studied, several recent papers have touched on related topics. In this Section, we compare our results to two studies with a similar setup.

While exploring the survival of molecular gas in outflows, \citet{Farber2022} performed simulations of cool ($10^3~\textrm{K}$), dusty clouds in hot winds. They modeled dust destruction in these simulations using fully dynamically coupled tracer particles, which they set to a ``dead" state if they encountered gas of temperature greater than $T_\mathrm{dest}$. They varied $T_\mathrm{dest}$ (from $10^4-10^6~\textrm{K}$) and the destruction time (from $0-4~t_\mathrm{cc}$), so that dust destruction depended on the duration of exposure to hot gas. Consistent with our results, they concluded that some dust could survive if it is shielded inside of large ($r_\mathrm{cl}\sim100~\textrm{pc}$) clouds, regardless of $T_\mathrm{dest}$, with higher survival fractions for larger clouds. However, because they implicitly assume short sputtering times, they generally found higher rates of dust destruction than in this paper.

\citet{Chen2023} carried out a similar study of the evolution of dusty molecular clouds in galactic winds. Like this work, they used a scalar-based dust model, with dust evolution subject to destruction through sputtering and growth through gas-phase metal accretion for a single grain size ($a=0.1~{\mu\textrm{m}}$). Accretion primarily occurred in $T<100~\textrm{K}$, dense gas, and had little effect beyond this regime. They came to a similar conclusion---that high fractions of dust survival for grains of this size are enabled by clouds that survive and entrain in galactic outflows. They found a typical ratio of $m_\mathrm{d}/m_\mathrm{d,i}=0.7$ for entrained clouds. This result is slightly lower than our findings for $0.1~{\mu\textrm{m}}$ grains given that it includes the effects of grain growth, although this may be a result of their somewhat longer simulation times. Because wind conditions are expected to change with distance from the galaxy to a regime less likely to result in sputtering, longer wind tunnel simulations may overestimate total dust destruction. They also found that clouds that fragmented more as they evolved exhibited higher levels of dust destruction, which is consistent with our results. Finally, they found that the dust-to-gas ratios of clouds decrease by roughly an order of magnitude as the cloud accretes pristine mixed-phase gas from the wind, similar to what we show in Figure \ref{fig:dtg}.

The inclusion of a cooling curve without a temperature floor may alter the distribution of dust within the cool phase, but we expect that it will not strongly affect overall dust survival. For example, although \citet{Chen2023} do allow cooling to lower temperatures, the cloud gas in their simulations is still largely in the warm ionized phase by volume, fully enveloping the smaller cold regions, and it is the evolution of the $\sim 10^{4}\,\mathrm{K}$ warm gas that governs cloud entrainment. Given this, overall cloud evolution is quite similar in simulations with and without a cooling floor since the warm phase always bridges the cloud-wind interface. Because dust destruction primarily occurs within and beyond this interface, we expect that our results for dust survival would be quite similar without a cooling floor.

\section{Conclusions} \label{sec:conclusions}

In this work, we have presented high-resolution simulations of cool, dusty clouds accelerated by hot galactic winds. These simulations were run using the Cholla hydrodynamics code, in which we implemented a new model to track the destruction of dust due to thermal sputtering. In particular, we conclude that:

\begin{enumerate}
    \item Cool clouds of gas can shield dust from sputtering in the hot phase of outflows, enabling a majority or total dust survival in some cases (Section~\ref{sec:results}). The overall amount of dust survival depends on how efficiently mixed-phase gas can re-accrete onto the cloud, which is controlled by the ratio $t_\mathrm{cool,minmix}/t_\mathrm{shear}$ (Section~\ref{sec:analytics}). We estimate that roughly $60\%$ of the $a=0.1~{\mu\textrm{m}}$ dust launched in clouds in outflows or starburst galaxies can survive for long enough to make it to the halo (Section~\ref{subsec:halo_dust}).
    
    \item Cloud shielding is not required for dust survival in outflows. Indeed, our simulations demonstrate that $a=0.1~{\mu\textrm{m}}$ grains can survive fully exposed to the hot phase, since the wind quickly carries dust away from the hottest, densest region of the wind nearest to the galaxy (Section~\ref{subsec:hot-phase-dust}).
    
    \item Hot gas in the CGM may be dust-enriched since large grains can survive in the wind. The contours of $t_\mathrm{sp}$ in Figure~\ref{fig:t_sp} illustrate that dust should be able to exist in these regions for long periods.
    
    \item In our simulations, grains below $a=0.1~{\mu\textrm{m}}$ are destroyed in significant amounts when exposed to the hot and sometimes even mixed phase of outflows. Only large clouds with $t_\mathrm{cool,minmix}/t_\mathrm{shear}\ll1$ enable small grains to travel significant distances (Section~\ref{subsec:grainsize}). This may indicate that shattering is a prominent source of small grains in the CGM.
    
    \item The dust-to-gas ratios of clouds decrease upwind as dust-free mixed-phase gas accretes onto their tails. We find that dust-to-gas ratios of clouds can decrease by roughly an order of magnitude between the time clouds are launched in outflows and when they reach the CGM (Section~\ref{subsec:dtg}).
\end{enumerate}

Taken together, our results demonstrate that significant fractions of dust can survive in outflows, either within clouds or within hot winds themselves. This provides a viable explanation for the large amounts of dust observed in CGM (and beyond) of nearby galaxies ~\citep[e.g.][]{Menard2010}. While dust survival in winds was able to be directly evaluated in our investigation owing to the use of very high mass and spatial resolution models, such resolution cannot be replicated within cosmological galaxy evolution models. Thus, to fully identify the implications of our study for dust/galaxy co-evolution~\citep[e.g.,][]{McKinnon2017}, our findings suggest that the treatment of dust growth and destruction in galaxy outflows needs to be revisited---possibly employing subgrid models---to more appropriately account for dust growth and destruction within otherwise unresolved multi-phase winds.

\begin{acknowledgments}
We thank the anonymous referee for their helpful review, which improved this paper. HMR thanks Robert Caddy, Alwin Mao, and Orlando Warren for many helpful discussions. This research was supported in part by the University of Pittsburgh Center for Research Computing, RRID:SCR\_022735, through the resources provided. Specifically, this work used the H2P cluster, which is supported by NSF award number OAC-2117681. This research also used resources of the Oak Ridge Leadership Computing Facility, which is a DOE Office of Science User Facility supported under Contract DE-AC05-00OR22725, using Frontier allocation AST181. E.E.S. acknowledges support from NASA TCAN grant 80NSSC21K0271, NASA ATP grant 80NSSC22K0720, StScI grant HST-AR-16633.001-A, and the David and Lucile Packard Foundation (grant no. 2022-74680).
P.T. acknowledges support from NASA ATP grant 80NSSC22K0716 and NSF AAG grant 2346977.
\end{acknowledgments}

\software{Cholla \citep{Schneider2015}, \texttt{numpy} \citep{VanDerWalt11}, \texttt{matplotlib} \citep{Hunter07},  \texttt{hdf5} \citep{hdf5} \texttt{seaborn} \citep{Waskom2021}}

\bibliography{dust-survival}{}
\bibliographystyle{aasjournal}

\appendix

\section{Additional simulations} \label{app:additional_sims}

\begin{table*}[t]
  {\centering
  \begin{tabular}{c c c c c c c c c c }
  \toprule
    Resolution & Dimensions~($r_\mathrm{cl}$) & $r_\mathrm{cl}~(\textrm{pc})$ & $a~(\mu\textrm{m})$ & $v_\mathrm{w}~(\textrm{km}/\textrm{s})$ & $T_\mathrm{w}~(\textrm{K})$ & $\chi$ & $m_\mathrm{d,sp}/m_\mathrm{d,i}$ & $t/t_\mathrm{sp,w}$ & $t_\mathrm{cool,minmix}/t_\mathrm{shear}$ \\
  \midrule

    $r_\mathrm{cl}$/16 & $64\times16\times16$ & 5 & 0.1 & $1\times10^3$ & $3\times10^7$ & $10^2$ & 0.14 & 0.24 & n/a \\
    
    $r_\mathrm{cl}/16$ & $64\times16\times16$ & 5 & 0.01 & $1\times10^3$ & $3\times10^7$ & $10^2$ & 0.81 & 1.9 & n/a \\

    $r_\mathrm{cl}$/16 & $64\times16\times16$ & 5 & 0.1 & $1\times10^3$ & $3\times10^6$ & $10^2$ & 0.06 & 0.14 & 8.4 \\

    $r_\mathrm{cl}$/16 & $64\times16\times16$ & 100 & 0.1 & $5\times10^2$ & $3\times10^7$ & $10^3$ & 0.24 & 7.2 & 0.83 \\
    
    $r_\mathrm{cl}$/16 & $64\times16\times16$ & 100 & 0.01 & $5\times10^2$ & $3\times10^7$ & $10^3$ & 0.76 & 59 & 0.83 \\

    $r_\mathrm{cl}$/16 & $64\times16\times16$ & 100 & 0.1 & $1\times10^3$ & $3\times10^6$ & $10^3$ & 0.03 & 4.0 & 0.42 \\
    
    $r_\mathrm{cl}$/16 & $64\times16\times16$ & 100 & 0.01 & $1\times10^3$ & $3\times10^6$ & $10^3$ & 0.26 & 36 & 0.42 \\
    
  \bottomrule
  \end{tabular}
  \caption{Additional simulations run in our parameter study. See Table~\ref{tab:sims} for a description of each column.}
  \label{tab:sims_app}}
\end{table*}

In Table~\ref{tab:sims_app}, we show the parameters for additional simulations we ran that are not explicitly discussed in Section~\ref{sec:results}. We find that these are broadly consistent with the results from the simulations we feature in Table~\ref{tab:sims}. One simulation of note is shown in the final row. This is a cloud in the survival regime with $0.01~{\mu\textrm{m}}$ dust. Here, we see a significant increase in the total amount of sputtering compared to our other $a=0.01~{\mu\textrm{m}}$ survived cloud simulation, even though the shear is only slightly higher (here, $t_\mathrm{cool,minmix}/t_\mathrm{shear}$ is higher by a factor of $\sim2$). Increasing the shear by another factor of two (shown two rows above this) at this grain size results in a majority of dust destruction. This illustrates that clouds must be well within the survival regime for small grain shielding to be efficient.

\section{Cloud tracking} \label{app:cloud_tracking}

As the wind accelerates the clouds in our simulations, they will be pushed down the simulation volume and eventually carried out of the simulation altogether. To avoid this, we implemented a cloud-tracking routine, allowing us to study the evolution of dust for longer. Our cloud tracking routine consists of a reference frame update which is applied to the entire grid after each time step. The mass-weighted average cloud velocity is given by

\begin{equation}
\langle v_x\rangle=\int_V\rho v_x dV,
\end{equation}

\noindent where $\rho$ and $v_x$ are the cloud densities and x-velocities. We integrate over the entire cloud (i.e. all gas that satisfies $\rho>\rho_\textrm{cl,i}/\chi^{1/2}$) to determine $\langle v_x \rangle$ and subtract the subsequent value from every cell in the grid to perform our reference frame update. This enables us to keep the majority of the cool gas mass in the simulation volume indefinitely.

We simulated our survived cloud (described in Table~\ref{tab:sims}) in a box size of $14\times2\times2~\textrm{kpc}^3$ for $\sim125~\textrm{Myr}$. We found that a total of $2\%$ of the initial dust mass was sputtered versus $1\%$ for the version of the simulation without cloud tracking. Thus, we conclude that our results are largely unchanged by the effect of cloud material leaving the box.

\section{Convergence} \label{app:convergence}

\begin{figure*}
\centering
\includegraphics[width=\textwidth]{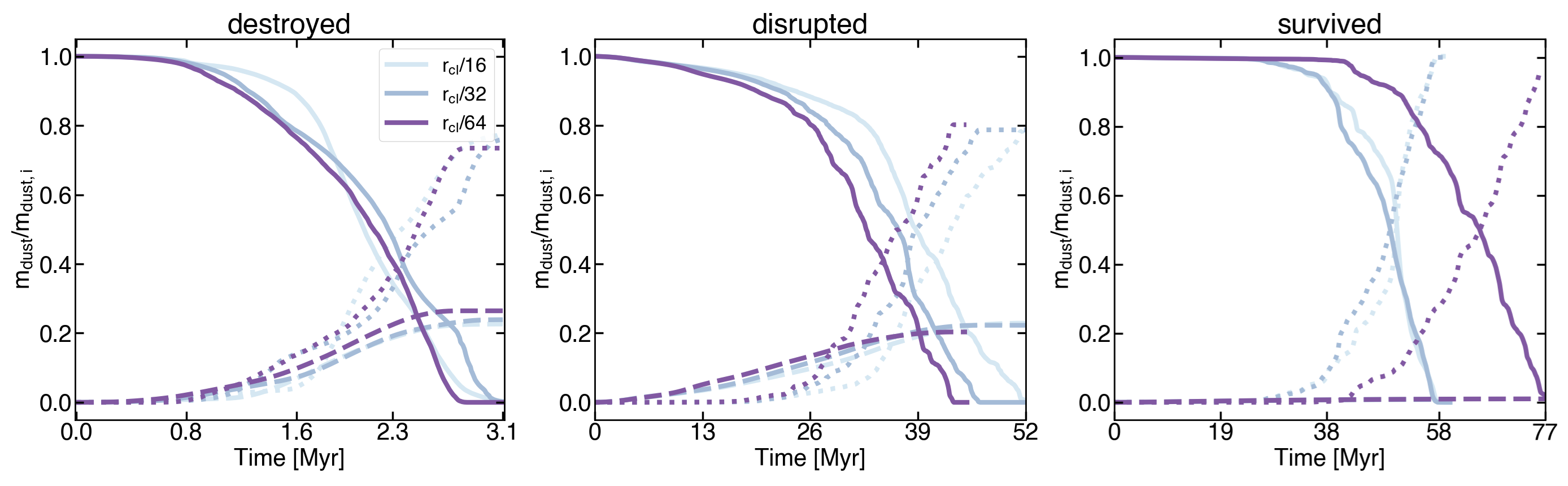}
\caption{Resolution study for the destroyed, disrupted, and survived cloud simulations (described in Table~\ref{tab:sims}). All runs use the same initial conditions (except the $r_\mathrm{cl}/64$ survived cloud, which was run in a box twice the length as the lower resolution runs). Note that some box sizes are slightly different than noted in Table~\ref{tab:sims} (namely, the destroyed and disrupted clouds use $100\times15\times15~r_\mathrm{cl}^3$ and $64\times16\times16~r_\mathrm{cl}^3$ boxes, respectively), but these box sizes are kept consistent between each resolution. The solid line shows the total amount of dust in the volume, the dotted line shows an estimate of how much dust has left the volume, and the dashed line is the total fraction of dust sputtered. Overall, the total amount of dust sputtered does not change with resolution.
\label{fig:res_comp}}
\end{figure*}

To test the resolution dependence of our results, we ran our destroyed, disrupted, and survived cloud simulations (described in Table~\ref{tab:sims} and in Section~\ref{sec:results}) for $a=0.1~{\mu\textrm{m}}$ grains at varying resolution. We repeated each simulation at $r_\mathrm{cl}/16$, $r_\mathrm{cl}/32$, and $r_\mathrm{cl}/64$. In Figure~\ref{fig:res_comp}, we compare the total amount of sputtering between each resolution. Overall, we find that our results for total dust sputtering are insensitive to resolution. For each regime of cloud evolution, essentially the same total amount of sputtering is observed between all three resolutions, as shown by the dashed lines in Figure~\ref{fig:res_comp}. In all cases, the primary difference between each resolution is the cloud dynamics, which are reflected in the total dust mass in the volume, shown in the solid (dust in volume) and dotted lines (dust that has exited the volume). This discrepancy arises because higher-resolution clouds are accelerated less efficiently in wind tunnel simulations \citep{Schneider2017}. Since cloud acceleration does not appear to affect dust sputtering, we run a majority of our simulations at $r_\mathrm{cl}/16$.

\section{Gas drag and nonthermal sputtering} \label{app:nonth}

To estimate the effect that nonthermal sputtering may have on dust survival in our simulations, we compare the gas drag and nonthermal sputtering times. Because nonthermal sputtering is a result of relative gas-grain velocities, it should be negligible in cases where the drag time (the time to accelerate dust grains to the gas velocity) is short compared to the nonthermal sputtering time. The definition of the drag time is given by

\begin{equation}
t_\mathrm{drag}\equiv \frac{2\sqrt{2}a\rho_\mathrm{gr}c_\mathrm{s}}{3n_\mathrm{H}k_\mathrm{B}TG(s)},
\end{equation}

\noindent where $a$ is the grain size, $\rho$ is the dust grain density, $c_\mathrm{s}$ is the sound speed, $n_\mathrm{H}$ and $T$ are the gas density and temperature, and $G(s)$ ($s\equiv v_\mathrm{rel}/(\sqrt{2}c_\mathrm{s})$) is a dimensionless term that quantifies the Coulomb drag (\citealt{Draine2011} and references therein). This can be approximated as

\begin{equation}
    t_\mathrm{drag}\approx0.59~\textrm{Myr}\,\Big(\frac{a}{\mu\textrm{m}}\Big)\Big(\frac{n_\mathrm{H}}{\textrm{cm}^{-3}}\Big)^{-1}\Big(\frac{T}{10^6~\textrm{K}}\Big)^{-1/2}G(\textrm{s})^{-1}
\end{equation}

\begin{figure}
\centering
\includegraphics[width=0.42\textwidth]{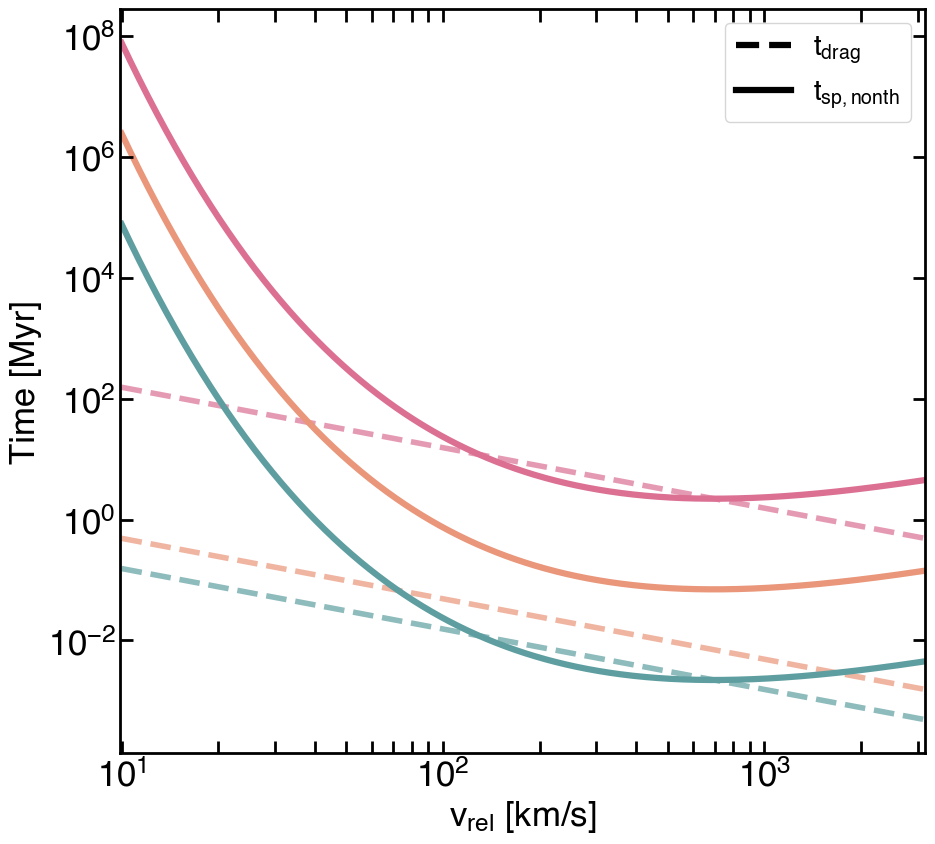}
\caption{Comparison between drag (solid line) and nonthermal sputtering (dashed line) times for the hot (pink), mixed (orange), and cool (green) phases of outflows for $a=0.1~{\mu\textrm{m}}$ silicate grains.
\label{fig:drag_nonth}}
\end{figure}

\noindent \citep{Hu2019}. In highly supersonic flows, $G(s)\propto s$, so we take $G(s)\approx s$ in order to obtain a rough estimate of the drag time, although we note that this approximation applies to neutral gas only. An approximation for ionized gas with $s\gtrsim1$ can be found in \citet{Hu2019}. Because $G(s)$ is in the denominator, we note that we are estimating an upper limit for the drag time by using this approximation. The nonthermal sputtering time is given by

\begin{equation}
    t_\mathrm{sp,nth}\approx0.33~\textrm{Myr}\Big(\frac{a}{\mu\textrm{m}}\Big)\Big(\frac{n_\mathrm{H}}{\textrm{cm}^{-3}}\Big)^{-1}\Big(\frac{Y_\mathrm{nth}}{10^{-6}{\mu\textrm{m}}\,\textrm{yr}^{-1}\,\textrm{cm}^3}\Big)^{-1}
\end{equation}

\noindent where $Y_\mathrm{nth}$ is the nonthermal sputtering yield, which depends on the dust-gas relative velocity (\citealt{Nozawa2006, Hu2019}). We compared these two timescales for $a=0.1~{\mu\textrm{m}}$ silicate grains (which have the highest sputtering yield) in cool, mixed, and hot outflow gas phases. The results are shown in Figure \ref{fig:drag_nonth}.

We compare the nonthermal sputtering and drag times as a function of gas-grain relative velocity in Figure \ref{fig:destroyed-snapshots}. Note that we use the cloud-wind relative velocity to characterize grain decoupling, but in actuality, the turbulent velocity in the mixing layer between the cloud and wind, $v_\mathrm{turb}$, is likely to dictate these mechanisms. Since $v_\mathrm{turb}$ is always smaller than the cloud-wind relative velocity \citep{Tan2024}, the cloud-wind relative velocity is effectively an upper limit $v_\mathrm{rel}$. Even in this extreme case, drag efficiently recouples dust and gas before nonthermal sputtering takes effect.

\begin{figure}
\centering
\includegraphics[width=0.42\textwidth]{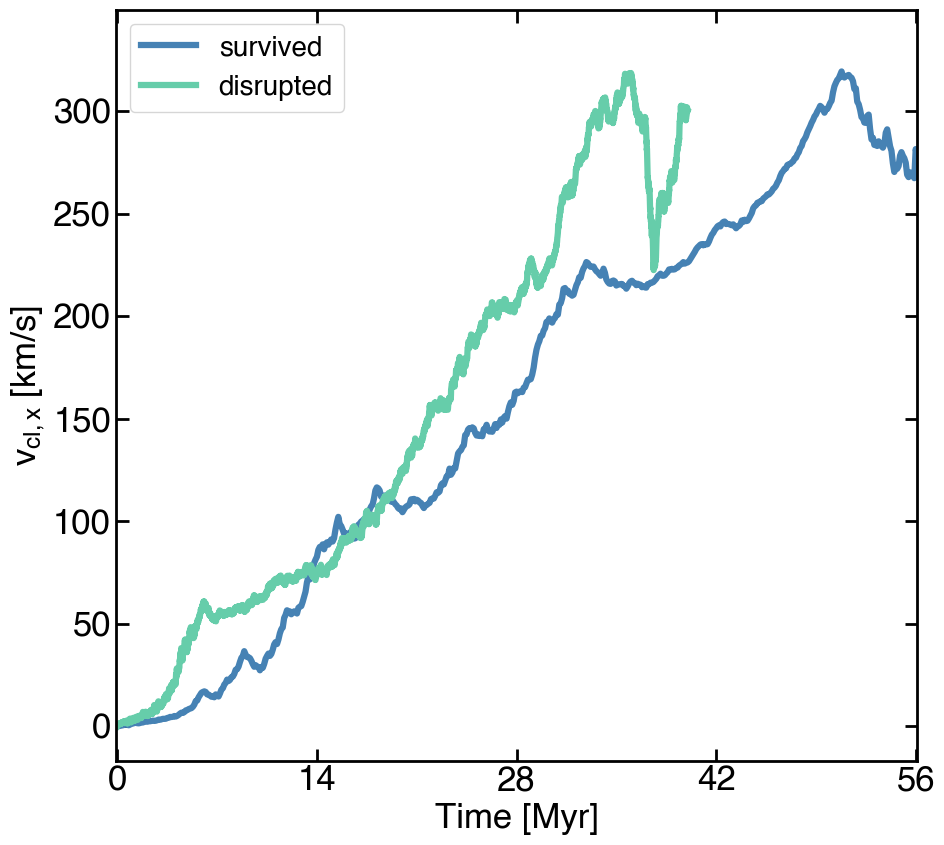}
\caption{The average cloud x-velocities as a function of time for our survived and disrupted cloud simulations (described in Table~\ref{tab:sims}).
\label{fig:v_cl}}
\end{figure}

For the mixed phase, drag timescales are always shorter than the nonthermal sputtering time. Near $v_\mathrm{rel}\sim300~{\textrm{km}\,\textrm{s}^{-1}}$, $t_\mathrm{sp,nth}$ is slightly shorter than $t_\mathrm{drag}$ for both hot and cool phases. Clouds in outflows can reach velocities comparable to this, but their acceleration happens gradually. Figure \ref{fig:v_cl} shows the evolution of average cloud velocity for the survived and disrupted clouds. In both cases, it takes tens of millions of years for the wind to accelerate the cloud to a velocity of $300~{\textrm{km}\,\textrm{s}^{-1}}$. This is because cloud acceleration is driven by mixing processes, which happen on relatively long timescales. The drag timescale for the cool and mixed phases is at least an order for magnitude shorter than the measured acceleration timescale for the cloud, so we conclude that gas and dust in these phases should remain dynamically coupled. The transfer of dust between phases is driven by mixing (dust cannot be instantaneously transferred to the hot phase with zero velocity), so relative velocities remain low enough that the drag force should take effect before nonthermal sputtering becomes significant, even in the hot phase. Since both timescales scale linearly with $a$, this conclusion applies for all grain sizes.

\end{document}